\documentclass{appolb}

\usepackage{epsf,rotate}
\usepackage{amsmath,amstext,amsbsy,amsopn,amsfonts,amssymb}
\usepackage[dvips]{graphicx}
\usepackage[latin1]{inputenc}
\usepackage[english]{babel}
\usepackage[T1]{fontenc}
\usepackage{psfrag}

\newcommand{\Log}{\mbox{Log}}

\newcommand{\eqdef}{\mbox{$ \stackrel{\rm def}{=}$}}

\newcommand{\asbar}{\ifmmode\wbar {\alpha}_{\rm s}\else{$\wbar{\alpha}_{\rm s}$}\fi}

 \catcode`\@=11

\def \lab #1 {\label{#1}}
\newcommand \ci [1] {\cite{#1}}

\newcommand\re[1]{(\ref{#1})}

\def \Im {\mathop{\rm Im}\nolimits}
\def \Re {\mathop{\rm Re}\nolimits}

\newcommand\lr[1]{{\left({#1}\right)}}
\newcommand \wbar [1] {\overline{#1}}
\newcommand \vev [1] {\langle{#1}\rangle}

\newcommand \ket [1] {|{#1}\rangle}

\newcommand \mybf[1] {\mbox{\boldmath$ {#1} $}}
\newcommand \ot[1] {\Hat{t}_{#1}}
\newcommand \otb[1] {\Hat{\wbar{t}}_{#1}}
\newcommand \oq[1] {\Hat{q}_{#1}}
\newcommand \oqb[1] {\Hat{\wbar{q}}_{#1}}
\def \SL  {\mbox{SL}}

\begin{document}
\eqsec
\title{Three particle Pomeron and odderon states in QCD
\author{{\bf Jan Kota{\'n}ski}
\address{M. Smoluchowski Institute of Physics, Jagellonian University\\
Reymonta 4, 30-059 Krak{\'o}w, Poland}}}


\maketitle

\begin{abstract}
{\normalsize
The scattering amplitude of  hadrons in high energy
Regge limit can be rewritten in terms of reggeized gluons, 
\ie Reggeons.
We consider three-Reggeon states that possess either $C=+1$
or $C=-1$ parity. In this work
using Janik-Wosiek method 
the spectrum of conformal charges 
is calculated for states with conformal Lorentz spin 
$n_h=0,1,2,3,\ldots\,$. Moreover  
corrections to WKB approximation are computed.
}
\end{abstract}


\vspace*{1cm}
\noindent TPJU-03/2006 \newline

\newpage

\bibliographystyle{unsrt}

\section{Introduction}

The reggeized gluon states, also called Reggeons, appear in 
scattering processes
of Quantum Chromodynamics (QCD) in
the Regge limit
where the square of the total energy $s$ is large while
the transfer of four-momentum $t$ is low and fixed.
In this limit the leading contribution to the scattering amplitude
of hadrons
is dominated by the exchange of 
intermediate particles,
Reggeons, which are the compound states 
of gluons 
\ci{gell,gell2,Gribov:1968fc,Fadin:1975cb,Bartels:1977hz,Cheng:1977gt}.

Even in
the simplifying Regge limit 
in the generalized leading logarithm approximation  
\ci{Bartels:1980pe,Kwiecinski:1980wb,Jaroszewicz:1980mq}
this problem is 
technically very complicated
due to the
non-abelian structure of QCD.
Many Reggeon wave-functions
are the eigenstates of a Hamiltonian which
is 
equivalent 
in the 't Hooft's multi-colour limit 
\ci{'tHooft:1973jz,Lipatov:1990zb,Lipatov:1993qn} 
to the Hamiltonian for the non-compact Heisenberg
$\SL(2,\mathbb{C})$ spin magnet.
Moreover, the multi-Reggeon system with $N$ particles 
is completely solvable \ci{Lipatov:1993yb,Faddeev:1994zg}.
Thus, it possesses a complete set of integrals of motion,
so called 
conformal charges $(q_2,\wbar q_2,q_3, \wbar q_3,\ldots,q_N,\wbar q_N)$
that commute with the Hamiltonian.
The eigenvalues of the $\SL(2,\mathbb{C})$ Hamiltonian are
also called the energies of the Reggeons.
The Schr\"odinger equation for the lowest non-trivial case,
\ie for $N=2$ Reggeons, was formulated and solved
by Balitsky, Fadin, Kuraev and Lipatov 
\ci{Balitsky:1978ic,Kuraev:1977fs,Fadin:1975cb}.
They calculated the energy of the Pomeron state
with $N=2$ Reggeons.
An integral equation for three and more Reggeons
was formulated in Refs.  
\ci{Bartels:1980pe,Kwiecinski:1980wb,Jaroszewicz:1980rw}
in 1980. 
However, it took almost twenty years to obtain
the solution for $N=3$, which corresponds also to the 
QCD odderon \ci{Janik:1998xj,Gauron:1987jt,Lukaszuk:1973nt}.
Finally, the solutions for higher $N=4,\ldots,8$
have been found recently in a series of papers 
\ci{Derkachov:2002pb,Korchemsky:2001nx,Derkachov:2002wz} 
written in collaboration with S.\'E. Derkachov,
G.P. Korchemsky and A.N. Manashov.
Similar approach was taken by de Vega and Lipatov 
\ci{DeVega:2001pu,deVega:2002im}.

Description of scattering amplitudes in terms of 
the reggeized gluon states is most frequently used for
the elastic 
scattering amplitude of two heavy hadrons whose masses are
comparable. Here the amplitude
is equal to a sum over the Regge poles. 
In the Regge limit, 
$s \to \infty$ and $t=\mbox{const}$
they give behaviour 
like $s^{\alpha(t)}$ where 
$\alpha(0)$ is called the intercept and its value
is close to $1$. 
On the other hand, the intercept is related to the
minimum of the Reggeon energy defined by the $\SL(2,\mathbb{C})$
Hamiltonian.
Thus, evaluating  the spectrum of the $\SL(2,\mathbb{C})$ 
XXX Heisenberg model
we can calculate the behaviour of the hadron scattering amplitudes.

Three reggeized gluon states in $t-$channel correspond
either to Pomeron exchanges with parity $C=+1$ or
to odderon states with $C=-1$.
The Janik-Wosiek solutions include both of them.
Using the duality symmetry as well as momentum representation
Bartels, Lipatov and Vacca found another solution with $q_3=0$
and $\alpha(0)=1$  
\ci{Lipatov:1998as,Bartels:1999yt,Bartels:2001hw,
Vacca:2000bk,Kovchegov:2003dm}.
On the other hand there was shown in 
Refs.\ \ci{Lipatov:1993yb,Faddeev:1994zg} that
in order to solve the problem and 
calculate the intercept one may use
$Q-$Baxter method \ci{Baxter}.
It was applied to three Reggeon problem in 
Refs.~\ci{Derkachov:2002pb,Korchemsky:2001nx,Derkachov:2002wz} 
and 
Refs.~\ci{DeVega:2001pu,deVega:2002im}, however only 
in the former ones there was a complete agreement with the Janik-Wosiek
method.
The intercept for the three-Reggeon function was also calculated
directly using variational method in Ref. \ci{Braun:1998mg}.
The results agree with values obtained by the exact $Q-$Baxter method.

In this work 
we present the generalized Janik-Wosiek method of constructing 
the Reggeon eigenstates. Moreover, we calculate
the reach spectrum of the the three Reggeon energy
and the spectrum of the conformal charges $\{q_3,\wbar q_3\}$
for $n_h=0,1,2,3$. At the end we evaluate corrections to the
WKB approximation \ci{Derkachov:2002pb}.

In Section 2 we discuss the current state of knowledge.
Next we describe construction of the Reggeon eigenfunctions
that consists in
solving the the differential eigenequations of the conformal charges.
We systematize the knowledge about the ansatzes for eigenstates
of $\{q_3,\wbar q_3\}$ 
for three reggeized gluon,
extend the calculations to an arbitrary complex spin $s$ and
derive differential eigenequations for the conformal charges $q_3$.
Moreover, we show solutions to the  differential eigenequations
with $N=3$ and $s=0$ and resum obtained series solutions for the $q_3=0$ case.
We construct an exact solution to the $\hat q_3-$eigenequation, 
which may be solved by a series method, and find
the quantization conditions for $\{q_3,\wbar q_3\}$ 
which come from single-valuedness of the Reggeon wave-function.
The numerical results of this method for $N=3$
are presented in 
Section 4.
They agree with solutions
found using $Q-$Baxter method 
\ci{Korchemsky:2001nx,Derkachov:2002wz}.
We present quantized values of $q_3$
for different Lorentz spins $n_h=0,\ldots,3$ 
as well as corrections to 
the WKB approximation. 
At the end we make final conclusions.

\section{Hamiltonian method}
\subsection{The formalism}

In the high energy Regge limit
\begin{equation}
s \to \infty  \qquad t={\rm const}
\lab{eq:rlim}
\end{equation}
the contribution to the scattering amplitude of two hadrons $A$ and $B$
from an exchange of three reggeized gluons
may be written as 
\begin{equation}
\mathcal{A}_{N=3}(s,t) = s \int d^2 z_0\,
\e^{i\vec z_0 \cdot \vec p}
\vev{\tilde{\Phi}_A(\vec z_0)|
\e^{- \wbar{\alpha}_s Y \tilde{\cal H}_3/4 }
\lr{{\vec{\partial}_1}^{\, 2} {{\vec{\partial}}_2}^{\, 2} {{\vec{\partial}}_3}^{\, 2} }^{-1}
 | \tilde{\Phi}_B(0)}\,,
\lab{eq:A}
\end{equation}
where $\partial_k=\partial/\partial z_k$, the rapidity $Y=\ln s/s_0$
and $\lr{\vec{\partial}_1^{\, 2} \vec{\partial}_2^{\, 2} \vec{\partial}_3^{\, 2}}^{-1}$ 
are gluon propagators.
Reggeized gluon states, which are also called Reggeons,
are effective particles. They interact with each other and propagate
in the $t-$channel.
The Hamiltonian $\widetilde{\cal H}_3$
is given \ci{Bartels:1980pe,Kwiecinski:1980wb,Jaroszewicz:1980mq} 
as a sum of $N=3$ BFKL kernels \ci{Balitsky:1978ic,Kuraev:1977fs,Fadin:1975cb}.
The wave-functions 
$\ket{\tilde \Phi_{A(B)}(\vec z_0)}\equiv \tilde 
\Phi_{A(B)}(\vec z_i-\vec z_0)$
describe the coupling of three Reggeons to the scattered particles.
The $\vec z_0 -$ integration fixes the momentum transfer $t=-{\vec p_{T}}^{\; 2}$ 
whereas the operators $1/{\vec \partial_k}^{\, 2}
$  
stand for two-dimensional 
transverse propagators.

Defining the functions
 $\Phi(\vec z)$ as
\begin{equation}
\widetilde\Phi(\vec z)=(-i)^3\partial_{z_1}
\partial_{z_2} \partial_{z_3}\Phi(\vec z)
\end{equation}
the scalar product in the amplitude (\ref{eq:A}) can be rewritten as
\begin{multline}
\vev{\tilde{\Phi}_A(\vec z_0)|
\e^{- \wbar{\alpha}_s Y  \tilde{\cal H}_3/4 }
\lr{{\vec \partial_1}^{\, 2} {\vec \partial_2}^{\, 2} {\vec \partial}_3^{\,2}}^{-1}
| \tilde{\Phi}_B(0)}=\\
=\vev{\Phi_A(\vec z_0)|
\e^{- \wbar{\alpha}_s Y {\cal H}^{(s=0,\wbar s=1)}_3/4 }
| \Phi_B(0)}\,.
\lab{eq:vev}
\end{multline}
The Hamiltonians, ${\cal H}_3$ and $\widetilde{\cal H}_3$ are
invariant under 
the coordinate transformation of the $\SL(2,\mathbb{C})$ group
\begin{equation}
z_{k}^{\prime }=\frac{az_{k}+b}{cz_{k}+d}\mbox {,}\qquad 
\wbar{z}_{k}^{\prime }=
\frac{\wbar{a}\wbar{z}_{k}+\wbar{b}}{\wbar{c}\wbar{z}_{k}+\wbar{d}}
\lab{eq:trcoords}
\end{equation}
with $k=1,2,3$ while  $ad-bc=\wbar{a}\wbar{d}-\wbar{b}\wbar{c}=1$
and they are related to each other as
\begin{equation}
{\cal H}^{(s=0,\wbar s=1)}_3=
\lr{\wbar \partial_1 \wbar \partial_2 \wbar \partial_3}\,
\widetilde{\cal H}_3\,
\lr{\wbar \partial_1 \wbar \partial_2 \wbar \partial_3}^{-1}\,.
\lab{eq:H-rel}
\end{equation}

Now, one may associate with each particle 
generators of transformation (\ref{eq:trcoords}) \ci{Derkachov:2001yn}.
This generators are a pair of mutually commuting holomorphic
and anti-holomorphic spin operators, 
$S_{\alpha }^{(k)}$ and $\wbar{S}_{\alpha }^{(k)}$.
They satisfy the standard commutation relations 
$\left[S_{\alpha }^{(k)},S_{\beta }^{(n)}\right]
=i\epsilon _{\alpha \beta \gamma }\delta ^{kn}S_{\gamma }^{(k)}$
and similarly for $\wbar{S}_{\alpha }^{(k)}$.
The generators act on the quantum
space of the $k$-th particle, $V^{(s,\wbar{s})}$ as 
differential operators 
\begin{eqnarray}
\nonumber
 S_{0}^{k}=z_{k}\partial _{z_{k}}+s\,, \quad  & 
S_{-}^{(k)}=-\partial _{z_{k}}\,, \quad  & 
S_{+}^{(k)}=z_{k}^{2}\partial _{z_{k}}+2sz_{k}\,,\\
 \wbar{S}_{0}^{k}=
\wbar{z}_{k}\partial _{\wbar{z}_{k}}+\wbar{s}\,, \quad  & 
\wbar{S}_{-}^{(k)}=-\partial _{\wbar{z}_{k}} \,, \quad  & 
\wbar{S}_{+}^{(k)}=
\wbar{z}_{k}^{2}\partial _{\wbar{z}_{k}}+2\wbar{s}\wbar{z}_{k}\,,
\lab{eq:spins}
\end{eqnarray}
where $ S_{\pm}^{(k)}=S_{1}^{(k)}\pm i S_{2}^{(k)}$
while the complex parameters, $s$ and $\wbar{s}$,
are called the complex spins. Thus, the Casimir operator reads 
\begin{equation}
\sum_{j=0}^2(S_j^{(k)})^{2}=(S_{0}^{(k)})^{2}+(S_{+}^{(k)}S_{-}^{(k)}
+S_{-}^{(k)}S_{+}^{(k)})/2=s(s-1)
\lab{eq:casimir}
\end{equation}
and similarly for the anti-holomorphic operator $(\wbar{S}^{(k)})^{2}$.
The eigenstates of the $\SL(2,\mathbb{C})$ invariant system transform
as \ci{CFT,Zuber:1995rj}
\begin{equation}
\Psi (z_{k},\wbar{z}_{k})\rightarrow \Psi ^{\prime }(z_{k},\wbar{z}_{k})
=(cz_{k}+d)^{-2s}(\wbar{c}\wbar{z}_{k}
+\wbar{d})^{-2\wbar{s}}\Psi 
(z_{k}^{\prime },
\wbar{z}_{k}^{\prime })\mbox {.}
\lab{eq:trpsi}
\end{equation}

Due to the invariance (\ref{eq:trcoords}) of the system
we can rewrite the Hamiltonian as
\begin{equation}
\mathcal{H}_{3}=H_{3}+\wbar{H}_{3}\mbox {,}\qquad 
[H_{3},\wbar{H}_{3}]=0\mbox { }
\lab{eq:sepH}
\end{equation}
in terms of the conformal spins (\ref{eq:spins})
\begin{equation}
H_{3}=\sum _{k=1}^{N}H(J_{k,k+1})\mbox {,}\qquad 
\wbar{H}_{3}=\sum _{k=1}^{3}H(\wbar{J}_{k,k+1})\mbox {,}
\lab{eq:Ham}
\end{equation}
where
\begin{equation}
H(J)=\psi (1-J)+\psi (J)-2\psi (1)
\lab{eq:HJ}
\end{equation}
 with $\psi (x)=d\ln \Gamma (x)/dx$ being the Euler function and
$J_{3,4}=J_{3,1}$. Here operators, $J_{k,k+1}$ and $\wbar{J}_{k,k+1}$,
are defined through the Casimir operators for the sum of the spins
of
the neighbouring Reggeons
\begin{equation}
J_{k,k+1}(J_{k,k+1}-1)=(S^{(k)}+S^{(k+1)})^{2}
\lab{eq:jvss}
\end{equation}
with $S_{\alpha }^{(4)}=S_{\alpha }^{(1)}$, and $\wbar{J}_{k,k+1}$ is
defined similarly. 

In statistical physics  (\ref{eq:Ham}) is called the Hamiltonian of 
the non-compact $\SL(2,\mathbb{C})$
XXX Heisenberg homogeneous spin magnets.
It describes the  nearest neighbour interaction
between three non-compact $\SL(2,\mathbb{C})$ spins attached to the particles
with periodic boundary conditions.

In QCD values of $(s,\wbar{s})$ depend on 
a chosen scalar product in the space of the wave-functions (\ref{eq:trpsi}) 
and they are usually equal to $(0,1)$ or $(0,0)$ 
\ci{Derkachov:2001yn,DeVega:2001pu}.

In order to find the high energy behaviour of the scattering amplitude 
we have to solve the Schr\"{o}dinger equation 
\begin{equation}
\mathcal{H}^{(s=0,\wbar s=1)}_{3}
\Psi (\vec{z}_{1},\vec{z}_{2},\vec{z}_{3})
=E_{N}\Psi (\vec{z}_{1},\vec{z}_{2},\vec{z}_{3})
\lab{eq:Schr}
\end{equation}
with the eigenstate $\Psi (\vec{z}_{1},\vec{z}_{2},\vec{z}_{3})$
being single-valued function on the planes $\vec{z}_k=(z_k,\wbar{z}_k)$,
normalizable with respect to the $\SL(2,\mathbb{C})$ invariant scalar
product\begin{equation}
||\Psi ||^{2}=\vev{\Psi|\Psi}=
\int d^{2}z_{1}d^{2}z_{2} d^{2}z_{3}
|\Psi (\vec{z}_{1},\vec{z}_{2},\vec{z}_{3})|^{2}\,,
\lab{eq:norm}
\end{equation}
where $d^{2}z_i=dx_idy_i=dz_id\wbar{z}_i/2$ with $\wbar z_i={z_i}^{\ast}$.

From the
point of view of the $\SL(2,\mathbb{C})$ spin chain
\begin{equation}
\widetilde{\cal H}_3=\mathcal{H}_3^{(s=0,\wbar s=0)}\,.
\lab{eq:HH}
\end{equation}

Indeed, the transformation $\wbar S_\alpha
\to\lr{\wbar\partial_1 \wbar\partial_2 \wbar\partial_3}\,\wbar S_\alpha\,
\lr{\wbar \partial_1 \wbar \partial_2 \wbar \partial_3}^{-1}$
maps the $\SL(2,\mathbb{C})$ generators of the spin $\wbar s=0$ 
into those with the spin
$\wbar s=1$. 

One concludes from \re{eq:norm} that the Hamiltonian 
${\cal H}^{(s=0,\wbar s=1)}_3$ is
advantageous with respect to 
$\widetilde{\cal H}_3={\cal H}^{(s=0,\wbar s=0)}_3$ 
as it has the quantum numbers
of the principal series representation of the $\SL(2,\mathbb{C})$ group.

However, one can also use $\mathcal{H}_3^{(s=0,\wbar s=0)}$ 
\ci{DeVega:2001pu} 
or even 
$\mathcal{H}_3^{(s=1,\wbar s=1)}$ \ci{deVega:2002im}. 
Then the factor $\lr{\partial_1 \partial_2 \partial_3}^{\mp 1}$
or $\lr{\wbar \partial_1 \wbar \partial_2 \wbar \partial_3}^{\mp 1}$ has 
to be included in the scalar product, \ie for $(s=1,\wbar s=1)$ we have 
\begin{equation}
||\Psi ||^{2}=
\int d^{2}z_{1}d^{2}z_{2}d^{2}z_{3}
|\lr{\wbar \partial_1 \wbar \partial_2 \wbar \partial_3}^{- 1}
\Psi (\vec{z}_{1},\vec{z}_{2},\vec{z}_{3})|^{2}\,.
\lab{eq:norml}
\end{equation}
Here
the scalar product is no longer in 
the principal series representation of the $\SL(2,\mathbb{C})$ group.
All these Hamiltonians with the corresponding scalar products are equivalent
up to the zero modes of the  
operators
$\lr{\partial_1\partial_2 \partial_3}^{\mp 1}$ and
$\lr{\wbar \partial_1 \wbar \partial_2  \wbar \partial_3}^{\mp 1}$.

A part of the amplitude (\ref{eq:A}) 
for $N=3$ reggeized gluons 
describes the leading contribution
of the states with parity $C=-1$, odderon, and subleading
contribution to the $C=+1$ states related to the Pomeron.
Both contributions are of the same order.

Instead solving the Hamiltonian eigenproblem (\ref{eq:Schr}) 
we can solve eigenproblems for conformal charges
\begin{eqnarray}
\displaystyle
\nonumber
\oq{2}&=&-2\sum _{i_{2}>i_{1}=1}^{N}
\left(\sum _{j_{1}=0}^{2}S_{j_{1}}^{(i_{1})}S_{j_{1}}^{(i_{2})}\right)\\
\nonumber
& = &\sum _{i_{2}>i_{1}=1}^{N}\left((z_{i_{2}i_{1}})^{2(1-s)}
\partial _{z_{i_{2}}}\partial _{z_{i_{1}}}
(z_{i_{2}i_{1}})^{2s}+2s(s-1)\right)\,,\\
\nonumber
\displaystyle
\oq{3} & = & 2\sum _{i_{1},i_{2},i_{3}=1}^{N}
\varepsilon _{i_{1}i_{2}i_{3}}
S_{0}^{(i_{1})}S_{1}^{(i_{2})}S_{2}^{(i_{3})}\\
& = & i^3 \sum _{i_{3}>i_{2}>i_{1}=1}^{N}
\nonumber
\left(z_{i_{1}i_{2}}z_{i_{2}i_{3}}z_{i_{3}i_{1}}
\partial _{z_{i_{3}}}\partial _{z_{i_{2}}}\partial _{z_{i_{1}}}
+sz_{i_{1}i_{2}}(z_{i_{2}i_{3}}-z_{i_{3}i_{1}})
\partial _{z_{i_{2}}}\partial _{z_{i_{1}}}
\right.\\
\nonumber
 &  &+  \left.sz_{i_{2}i_{3}}(z_{i_{3}i_{1}}-z_{i_{1}i_{2}})
\partial _{z_{i_{3}}}\partial _{z_{i_{2}}} 
+ sz_{i_{3}i_{1}}(z_{i_{3}i_{1}}-z_{i_{1}i_{2}})
\partial _{z_{i_{3}}}\partial _{z_{i_{2}}} \right.\\
& & - \left.  2s^{2}z_{i_{1}i_{2}}\partial _{z_{i_{3}}}
-2s^{2}z_{i_{2}i_{3}}\partial _{z_{i_{1}}}
-2s^{2}z_{i_{3}i_{1}}\partial _{z_{i_{2}}}\right)\,,
\lab{eq:q2q3}
\end{eqnarray}
where $z_{ij}=z_{i}-z_{j}$. Similar relations hold for 
the anti-holomorphic sector.
This gives
for the $\SL(2,\mathbb{C})$ spin $s=0$
\begin{equation}
\oq{3}= - i
z_{12}z_{23}z_{31}\partial_{z_1}
\partial_{z_2}\partial_{z_3}
\lab{eq:qks0}
\end{equation}
and for $s=1$
\begin{equation}
\oq{3}=- i
\partial_{z_1}\partial_{z_2}\partial_{z_3}
z_{12} z_{23}z_{31}\,.
\lab{eq:qks1}
\end{equation}

The eigenvalues of the quadratic conformal charge read:
\begin{equation}
q_2=-h(h-1)+3 s(s-1) \quad 
\bar q_2=-\bar h(\bar h-1)+3 \bar s(\bar s-1) \quad 
\lab{eq:q2}
\end{equation}
where conformal weights satisfy
\begin{equation}
h=\frac{1+n_{h}}{2}+i\nu _{h}\mbox {,}\qquad \wbar h=
\frac{1-n_{h}}{2}+i\nu _{h},
\lab{eq:hpar}
\end{equation}
while the eigenvalues of the cubic operator $\bar q_3= q_3^*$.
The parameter $n_{h}$ has the meaning of the two-dimensional
Lorentz spin of the particle, whereas $\nu_{h}$ defines its
scaling dimension.
They define the transformation law of the Hamiltonian eigenstates
\begin{multline}
\Psi (\vec{z}_{1'0'},\vec{z}_{2'0'},\vec{z}_{3'0'})
=(cz_{0}+d)^{2h}(\wbar{c}\wbar{z}_{0}+\wbar{d})^{2\wbar{h}} \\
\times  \left(\prod _{k=1}^{3}(cz_{k}+d)^{2s_{k}}(\wbar{c}\wbar{z}_{k}
+\wbar{d})^{2\wbar{s}_{k}}
\right)
\Psi (\vec{z}_{10},\vec{z}_{20},\vec{z}_{30})
\lab{eq:trpsiz0}
\end{multline}
with
\begin{equation}
\Psi _{\vec{p}}(\vec{z}_{1},\vec{z}_{2},\vec{z}_{3})=
\int d^{2}z_{0}e^{i\vec{z}_{0}\cdot \vec{p}}
\Psi (\vec{z}_{1}-\vec{z}_{0},\vec{z}_{2}-\vec{z}_{0},
\vec{z}_{3}-\vec{z}_{0})\,.
\lab{eq:Psip}
\end{equation}

The eigenstates $\Psi (\vec{z}_{10},\vec{z}_{20},\vec{z}_{30})$
belonging to $V^{(h,\wbar{h})}$ are labelled by the centre-of-mass coordinate
$\vec{z}_{0}$ and can be chosen to have the $\SL(2,\mathbb{C})$
transformation properties
with $z_{0}$ and $\wbar{z}_{0}$ transforming in the same way as $z_{k}$
and $\wbar{z}_{k}$, (\ref{eq:trcoords}).

Conformal charges commute also with cyclic particle permutation 
operator
\begin{equation}
 \mathbb{P}\Psi _{q,\wbar q}(\vec{z}_{1},\vec{z}_{2},\vec{z}_{3})  
\eqdef  \Psi _{q,\wbar q}(\vec{z}_{2},\vec{z}_{3},\vec{z}_{1})
= e^{i\theta_3 (q,\wbar q)}\Psi _{q,\wbar q}(\vec{z}_{1},\vec{z}_{2},\vec{z}_{3})\,,
\lab{eq:Psym}
\end{equation}
where the conformal charges are denoted by $q\equiv(q_2,q_3)$ 
and $\wbar q\equiv(\wbar q_2,\wbar q_3)$.

The phase $\theta_3 (q)$ which is connected with eigenvalues of $\mathbb{P}$ 
is called 
the quasimomentum. It takes the following values 
\ci{Derkachov:2002pb} 
\begin{equation}
\theta_3(q,\wbar{q})=2 \pi \frac{k}{3}, \qquad \mbox{for } k=0,1,\ldots,2\,.
\lab{eq:quask}
\end{equation}
The eigenstates of the conformal
charges $\oq{k}$ diagonalize ${\cal H}$ and $\mathbb{P}$. 

Let us also define another operator
\begin{equation}
 \mathbb{M}\Psi ^{\pm }(\vec{z}_{1},\vec{z}_{2},\vec{z}_{3})  
\eqdef  \Psi ^{\pm }(\vec{z}_{3},\vec{z}_{2},\vec{z}_{1}) 
= \pm \Psi ^{\pm }(\vec{z}_{1},\vec{z}_{2},\vec{z}_{3}),
\lab{eq:Msym}
\end{equation}
so called mirror permutation operator
which has two eigenvalues, $\pm 1$.
The $\mathbb{M}$ operator commutes with the Hamiltonian but
it does not commute with $\hat q_2$ and $\bar q_3$.

The cyclic and mirror permutation symmetries come
from the Bose symmetry.
Physical states should possess both symmetries. 

It turns out that adding colour factor
corresponding to the antisymmetric constant $f_{abc}$
for the odd-mirror states and the symmetric one $d_{abc}$
for the even-mirror states, we are able to restore Bose symmetry.
The tensor $f_{abc}$ corresponds to the Pomeron states while
$d_{abc}$ is related to the odderon states. 
Therefore, in order to check
a $C-$parity of a given state we need to study its parity under 
the mirror permutation $\mathbb{M}$. 
The states $\Psi$ satisfying $\mathbb{M}\, \Psi=-\Psi$ are the Pomeron states
whereas states for which $\mathbb{M}\, \Psi=+\Psi$ are the odderon states.

\section{Various ansatzes}

Using the eigenequation for $\hat q_2$ we get Lipatov's ansatz 
\ci{Lipatov:1998as}
for holomorphic part of the eigenstates:
\begin{equation}
 \Psi (z_{10},z_{20},z_{30})   =  
\frac{1}{(z_{10} z_{20} z_{30})^{2s}}
\left(\frac{z_{31}}{z_{10}z_{30}}\right)^{h-3 s}F(x)\,,
\lab{eq:orgAn}
\end{equation}
where  $x=\frac{z_{12} z_{30}}{z_{10} z_{32}}$. If we substitute 
$F(x)=G(x) \left(\frac{(x-1)^2}{-x}\right)^{h/3-s}$ 
we get another ansatz
\begin{equation}
\Psi (z_{10},z_{20},z_{30})
=\frac{1}{(z_{12}z_{23}z_{31})^{s}}
\left(\frac{z_{31}z_{12}z_{23}}{(z_{10})^{2}
(z_{20})^{2}(z_{30})^{2}}\right)^{\frac{h}{3}}G(x)\,.
\lab{eq:symAn}
\end{equation}

As we can see we obtained a totally symmetric
ansatz which is equivalent to the original one.

Both ansatzes have advantages and disadvantages. The symmetric one 
(\ref{eq:symAn})
is appropriate if we want to deal with the particle symmetries.
The original one (\ref{eq:orgAn})
has a simpler structure, it contains powers of
$h$ (not $h/3$), so we can use it when we want to construct proper
single-valuedness conditions. 
One can easily notice that $z^{h}\wbar{z}^{\wbar{h}}$
is single-valued because $h-\wbar{h}=n_{h}\in \mathbb{Z}$.

\subsection{Derivation of differential equations for $N=3$}

From the $\hat q_3-$eigenequation we get the third order differential equation
\begin{multline}
i q_{3}F(x)=
(3s-h)(h-1-s)(1-2x)F(x)+(((h-2)(h-1)(x-1)x\\
+s^{2}(2+11(x-1)x)+s(2-2h(1-2x)^{2}+11(x-1)x))F^{\prime}(x)\\
+(2+h-3s)(1-x)x(2x-1)F^{\prime\prime}(x)+(x-1)^{2}x^{2}F^{(3)}(x)\,,
\lab{eq:qF}
\end{multline}
where $q_3$ is a complex eigenvalue of $\hat{q}_3$.
From the QCD point of view the most interesting cases are for $s=0$:
\begin{multline}
i q_{3}F(x)=
(h-1)(h-2)x(x-1)F^{\prime}(x)\\
+(h-2)(x-1)x(1-2x)F^{\prime\prime}(x)
+x^{2}(x-1)^{2}F^{(3)}(x)
\lab{eq:ees0z32}
\end{multline}
and for $s=1$
\begin{multline}
i q_{3}F(x)=
(h-3)(h-2)(2x-1)F(x)+\\
+((4-2h-(h-8)(h-3)x+(h-8)(h-3)x^{2})F^{\prime}(x)\\
+(5-h)(x-1)x(2x-1)F^{\prime\prime}(x)
+(x-1)^{2}x^{2}F^{(3)}(x)\,.
\lab{eq:qFs1}
\end{multline}
The first such solution was derived and found numerically in 
\ci{Janik:1998xj}.

We solve the above equations by the series method
choosing representation $(s,\bar s)=(0,0)$ which gives the same
equation in holomorphic and antiholomorphic sectors.
The differential equation (\ref{eq:ees0z32})
has three regular singular points at $x=0$, $x=1$ and $x=\infty$.
To
generate solutions around other singular points we exchange variables. For
the case $x=1$ we can use a substitution $x=1-y$:
\begin{multline}
i q_{3}G(y)=
(h-1)(h-2)y(1-y)G^{\prime}(y)\\
+(h-2)(1-y)y(1-2y)G^{\prime\prime}(y)
-y^{2}(y-1)^{2}G^{(3)}(y).
\lab{eq:ees0z32x1m}
\end{multline}
with $F(1-y)=G(y)$
and for $x=\infty$ we have $x=1/y$:
\begin{multline}
q_{3}G(y)=i\left((y-1)(h+1)(h-2y)G^{\prime}(y)
\right. \\ \left.
+(y-1)y(2(h+1)-(h+4)y) G^{\prime\prime}(y) 
+y(y-1)^{2}G^{(3)}(y)\right)\,.
\lab{eq:ees0z32xi}
\end{multline}
where $F(1/y)=G(y)$.

In Ref. \ci{Janik:1998xj} Eq. (\ref{eq:ees0z32})
was solved for the case with $h=\wbar h =1/2$ and $\Re[q_3]=0$.
Next in 
Refs.~\ci{Korchemsky:1999is,Praszalowicz:1998pz,Kotanski:2001iq}
states with $n_h=0,2$ and $\nu_h \in \mathbb{R}$
as well as  $n_h=0$ and $i \nu_h \in \mathbb{R}$
were found. In this work we extend these calculations to 
$n_h=0,1,2,3$ and $\nu_h \in \mathbb{R}$.
To get a broader perspective on the problem
the reader is referred to Ref.~\ci{Kotanski:2005ci}.

\subsection{Wave-function for $s=\wbar{s}=0$ around $x=\wbar{x}=0$}

In order to obtain the full-complex solution 
for three-Reggeon state containing
the holomorphic and anti-holomorphic parts we have to glue together
solutions in these sectors:
\begin{equation}
\Phi_{q,\wbar{q}}(\{ z_{i}\},\{\wbar{z}_{i}\})=
\wbar{u}_{\wbar{q}}(\{\wbar{z}_{i}\})^{T}\cdot 
A^{(0)}(h,\wbar{h},q_{3},\wbar{q}_{3})\cdot u_{q}(\{ z_{i}\})\,,
\lab{eq:Psizz}
\end{equation}
where we 
use a ($3\times 3$) 
mixing-matrix, 
$A_{q,\wbar{q}}^{(0)}$, \ci{Janik:1998xj}
which does not depend on particle coordinates but only
on $q\equiv\{ q_{2},q_{3}\}$.
From the QCD point of view we have two possibilities of gluing solutions:
$(s=0,\wbar{s}=0)$ \ci{DeVega:2001pu} 
and $(s=0,\wbar{s}=1)$ \ci{Derkachov:2001yn}.
These two cases are equivalent except 
for the zero modes of the highest conformal
charge $\hat{q}_{N}$. Let us consider the first case, $s=\wbar{s}=0$.

The antiholomorphic and holomorphic
conformal charges are related by conditions $\wbar{h}=1-h^{*}$
and $\wbar{q}_{k}=q_{k}^{*}$. The wave-function has to be single-valued.
This condition defines the structure of the mixing-matrix. 

For $h\not\in\mathbb{Z}$ and $\hat{q}_{3}\ne0$ we have solutions
of the following type
\begin{eqnarray}
\nonumber
u_{1}(x) & = & x^{h}\sum_{n=0}^{\infty}a_{n,r_{1}}x^{n}\,,\\
\nonumber
u_{2}(x) & = & x^{1}\sum_{n=0}^{\infty}a_{n,r_{2}}x^{n}\,,\\
u_{3}(x) & = & x^{0}\sum_{n=0}^{\infty}b_{n,r_{3}}x^{n}+x^{1}
\sum_{n=0}^{\infty}a_{n,r_{2}}x^{n}\mbox{Log}(x)
\lab{eq:u-zero}
\end{eqnarray}
and similarly in the anti-holomorphic sector. 
The recurrence relations for the $a_{n,r_i}$ 
are given in Appendix C.
One can notice that $x^{a}\wbar{x}^{b}$
is single-valued only if $a-b\in\mathbb{Z}$. Moreover we have also
terms with $\mbox{Log}(x)$ which have to give in a sum
 $\mbox{Log}(x\wbar{x})$.
So in this case we have a mixing matrix of the form
\begin{equation}
A^{(0)}(h,\wbar{h},q_{3},\wbar{q}_{3})=
\left[
\begin{array}{ccc}
\alpha & 0 & 0\\
0 & \beta & \gamma\\
0 & \gamma & 0
\end{array}
\right]\,,
\end{equation}
where $\alpha$, $\beta$, $\gamma$ are arbitrary.
In the above matrix we have $A_{12}=A_{13}=A_{21}=A_{31}=0$ in order
to eliminate multi-valuedness coming from the power-terms, $A_{23}=A_{32}$
to obtain single-valuedness in $\Log(x)-$terms and $A_{33}=0$ because the term
$\mbox{Log}(x)\mbox{Log}(\wbar{x})$ is not single-valued on
the $\vec{x}-$plane.

In the case of $q_{3}=0$ and $h\not\in\{0,1\}$ we don't have any 
$\Log(x)-$terms
so the mixing matrix looks like 
\footnote{{\it Greek} variables in each  $A-$matrix have different 
numerical values}
\begin{equation}
A^{(0)}(h,\wbar{h},q_{3}=0,\wbar{q}_{3}=0)=
\left[
\begin{array}{ccc}
\beta & 0 & 0\\
0 & \alpha & \rho\\
0 & \varepsilon & \gamma
\end{array}
\right]\,.
\end{equation}

For $q_{3}\ne0$ and $h\in\mathbb{Z}$ we have a solution with
only integer powers
of $x$ and solutions without a Log, with one-Log and with a double-Log.
The structure of the matrix is
\begin{equation}
A^{(0)}(h\in\mathbb{Z},
\wbar{h}\in\mathbb{Z}
,q_{3},\wbar{q}_{3})
=\left[\begin{array}{ccc}
\alpha & \beta & \gamma\\
\beta & 2\gamma & 0\\
\gamma & 0 & 0
\end{array}\right]\,.
\end{equation}

In the last case for $q_{3}=0$ and $h\in\{0,1\}$ we have only integer powers
of $x$ and the third solution with one-Log term. The matrix 
has a form
\begin{equation}
A^{(0)}(h\in\{0,1\},
\wbar{h}\in\{0,1\}
,q_{3}=0,\wbar{q}_{3}=0)=
\left[
\begin{array}{ccc}
\alpha & \gamma & 0\\
\beta & \rho & \varepsilon\\
0 & \varepsilon & 0
\end{array}
\right]\,.
\end{equation}

\subsection{Wave-function for $s=\wbar{s}=0$ around $x=\wbar{x}=1^{-}$
and $x=\wbar{x}=\infty^-$}

We construct the wave-function 
around the other singular point 
exactly in the same way, obtaining matrices 
$A^{(1^{-})}(h,\wbar{h},q_{3},\wbar{q}_{3})$,
$A^{(1^{+})}(h,\wbar{h},q_{3},\wbar{q}_{3})$ (around $1$) and 
$A^{(\infty^{-})}(h,\wbar{h},q_{3},\wbar{q}_{3})$ (around $\infty$).

Thus, we have the wave-functions similar to
(\ref{eq:Psizz}).
For $h\not\in\mathbb{Z}$ and $\hat{q}_{3}\ne0$  get
\begin{eqnarray}
\lab{eq:u-one}
u_{1}(x) & = & (1-x)^{h}\sum_{n=0}^{\infty}a_{n,r_{1}}(1-x)^{n}\,,\\
\nonumber
u_{2}(x) & = & (1-x)^{1}\sum_{n=0}^{\infty}a_{n,r_{2}}(1-x)^{n}\,,\\
\nonumber
u_{3}(x) & = & (1-x)^{0}\sum_{n=0}^{\infty}b_{n,r_{3}}(1-x)^{n}
+(1-x)^{1}\sum_{n=0}^{\infty}a_{n,r_{2}}(1-x)^{n}\mbox{Log}(1-x)
\end{eqnarray}
and similarly in the anti-holomorphic sector. 
The mixing matrices take the following forms

\begin{equation}
A^{(1)}(h,\wbar{h},q_{3},\wbar{q}_{3})=\left[\begin{array}{ccc}
\alpha & 0 & 0\\
0 & \beta & \gamma\\
0 & \gamma & 0\end{array}\right]\,,
\quad 
\end{equation}
\begin{equation}
A^{(1)}(h,\wbar{h},q_{3}=0,\wbar{q}_{3}=0)=
\left[
\begin{array}{ccc}
\beta & 0 & 0\\
0 & \alpha & \rho\\
0 & \varepsilon & \gamma
\end{array}
\right]\,,
\end{equation}

\begin{equation}
A^{(1)}(h\in\mathbb{Z},
\wbar{h}\in\mathbb{Z}
,q_{3},\wbar{q}_{3})
=\left[
\begin{array}{ccc}
\alpha & \beta & \gamma\\
\beta & 2\gamma & 0\\
\gamma & 0 & 0
\end{array}
\right]\,,
\end{equation}

\begin{equation}
A^{(1)}(h\in\{0,1\},
\wbar{h}\in\{0,1\}
,q_{3}=0,\wbar{q}_{3}=0)
=\left[\begin{array}{ccc}
\alpha & \gamma & 0\\
\beta & \rho & \varepsilon\\
0 & \varepsilon & 0\end{array}\right]\,.
\end{equation}

Similarly, we proceed around $x=\infty^-$.
For $h\not\in\mathbb{Z}$ and $\hat{q}_{3}\ne0$ we have solutions
of type
\begin{eqnarray}
\lab{eq:u-inf}
u_{1}(x) & = & (1/x)^{0}\sum_{n=0}^{\infty}a_{n,r_{1}}(1/x)^{n}\,,\\
\nonumber
u_{2}(x) & = & (1/x)^{1-h}\sum_{n=0}^{\infty}a_{n,r_{2}}(1/x)^{n}\,,\\
\nonumber
u_{3}(x) & = & (1/x)^{-h}\sum_{n=0}^{\infty}b_{n,r_{3}}x^{n}
+(1/x)^{1-h}\sum_{n=0}^{\infty}a_{n,r_{2}}x^{n}\mbox{Log}(x)
\end{eqnarray}
and similarly in the anti-holomorphic region.
In this case we have the matrices 
\begin{equation}
A^{(\infty)}(h,\wbar{h},q_{3},\wbar{q}_{3})=\left[\begin{array}{ccc}
\alpha & 0 & 0\\
0 & \beta & \gamma\\
0 & \gamma & 0\end{array}\right]\,,
\quad
\end{equation}
\begin{equation}
A^{(\infty)}(h,\wbar{h},q_{3}=0,\wbar{q}_{3}=0)=\left[\begin{array}{ccc}
\beta & 0 & 0\\
0 & \alpha & \rho\\
0 & \varepsilon & \gamma\end{array}\right]\,,
\end{equation}

\begin{equation}
A^{(\infty)}(h\in\mathbb{Z},
\wbar{h}\in\mathbb{Z}
,q_{3},\wbar{q}_{3})=
\left[
\begin{array}{ccc}
\alpha & \beta & \gamma\\
\beta & 2\gamma & 0\\
\gamma & 0 & 0
\end{array}
\right]\,,
\end{equation}

\begin{equation}
A^{(\infty)}(h\in\{0,1\},
\wbar{h}\in\{0,1\}
,q_{3}=0,\wbar{q}_{3}=0)=\left[\begin{array}{ccc}
\alpha & \gamma & 0\\
\beta & \rho & \varepsilon\\
0 & \varepsilon & 0\end{array}\right]\,.
\end{equation}

\subsection{Transition matrices between solutions around different poles}

The above solutions around $x=0,1,\infty$ 
have a convergence radius equal to the 
difference between the two singular points:
the point around which the solution is defined and the nearest
one of the remaining two.
In order to define a global solution which is convergent
in the entire complex plane we have to glue
the solutions defined around different singular points. 
This can be done by expanding one set of solutions 
in terms of the other solutions in the overlap region of
the two considered solutions.
Thus, in the overlap region we can define the 
transition matrices $\varDelta$, $\varGamma$, where
\begin{eqnarray}
\nonumber
\vec{u}^{(0)}(x,q) &=&\varDelta(q) \vec{u}^{(1)}(x,q)\,, \\
\vec{u}^{(1)}(x,q) &=&\varGamma(q) \vec{u}^{(\infty)}(x,q).
\lab{eq:u01}
\end{eqnarray}

Matrices, $\varDelta$ and  $\varGamma$, are constructed in terms of the ratios
of certain determinants \ci{Wosiek:1996bf}.
For example, to calculate the matrix $\varDelta$ we construct a 
Wro\'nskian
\begin{equation}
W=
\begin{vmatrix}
u_1^{(1)}(x;q)& u_2^{(1)}(x;q)& u_3^{(1)}(x;q) \\
{u'}_1^{(1)}(x;q)& {u'}_2^{(1)}(x;q)& {u'}_3^{(1)}(x;q) \\ 
{u''}_1^{(1)}(x;q)& {u''}_2^{(1)}(x;q)& {u''}_3^{(1)}(x;q) 
\end{vmatrix}.
\lab{eq:wron}
\end{equation}
Next we construct determinants $W_{ij}$ which are obtained from $W$ 
by replacing 
$j$-th column by the $i$-th solution around $x=0$, \ie for 
$i=1$ and $j=2$ we have
\begin{equation}
W_{12}=
\begin{vmatrix}
u_1^{(1)}(x;q_3)& u_1^{(0)}(x;q_3)& u_3^{(1)}(x;q_3) \\
{u'}_1^{(1)}(x;q_3)& {u'}_1^{(0)}(x;q_3)& {u'}_3^{(1)}(x;q_3) \\ 
{u''}_1^{(1)}(x;q_3)& {u''}_1^{(0)}(x;q_3)& {u''}_3^{(1)}(x;q_3) 
\end{vmatrix}.
\end{equation}
The matrix elements $\varDelta_{ij}$ 
are given by
\begin{equation}
\varDelta_{ij}=\frac{W_{ij}}{W}.
\lab{eq:Dij}
\end{equation}
Matrix 
$\varDelta$ 
does not depend on $x$, but only on $q_k$.
In the similar way we can get matrices
$\varGamma$ 
and their anti-holomorphic equivalents:
$\overline{\varDelta}$, $\overline{\varGamma}$.

Substituting  equation (\ref{eq:u01}) into the wave-function
(\ref{eq:Psizz}), one finds the following conditions for
continuity of the matrix $A(\overline{q},q)$:
\begin{eqnarray}
\overline{\varDelta}^T(\overline{q}_3)A^{(0)}(\overline{q}_3,q_3)
\varDelta(q_3)&=
A^{(1)}(\overline{q}_3,q_3)\,, 
\lab{eq:azao}
\\
\overline{\varGamma}^T(\overline{q}_3)A^{(1)}(\overline{q}_3,q_3)
\varGamma(q_3)&=
A^{(\infty)}(\overline{q}_3,q_3)\,.
\lab{eq:aoai}
\end{eqnarray}
Each Equation, (\ref{eq:azao})-(\ref{eq:aoai}), 
consists of nine equations.   
Solving them numerically, we obtain values of parameters 
$\alpha,\beta,\gamma,\ldots$
as well as quantized values of the conformal charges, $q_k$ and 
$\wbar{q}_k$. 
We have verified numerically that the spectrum
of $q_k$ obtained by the above method is equivalent 
to the spectrum obtained using the 
Baxter $Q-$operator method which is presented in next Section.

\subsection{Additional conditions coming from the particle permutation 
invariance}

Multi-Reggeon states have additionally the cyclic (\ref{eq:Psym}) 
and mirror permutation
(\ref{eq:Msym}) symmetries. 
The conformal charges commute only with $\mathbb{P}$.
Thus, the eigenstates 
of $\oq{2}$ and $\oq{3}$ (\ref{eq:q2q3})
are usually not eigenstates of $\mathbb{M}$,
so they usually have mixed $C$-parity. 
Therefore in order to get solutions of a given $C$-parity
we have to impose mirror symmetry. 
Let us illustrate this for the case of $q_3=0$.

For this case
we can easily resum the series solutions, see Appendix C.
Let us take the case for $h\not\in\{0,1\}$.
The eigenequation
for the cyclic permutation $\mathbb{P}$ with quasimomentum $\theta_3(q)$ gives 
the following condition
\begin{multline}
w^{h}\wbar{w}^{\wbar{h}}\left(\beta
+\gamma(-x)^{h}(-\wbar{x})^{\wbar{h}}+\alpha(x-1)^{h}(\wbar{x}-1)^{\wbar{h}}
\right. \\ \left.
+\rho(-x)^{h}(\wbar{x}-1)^{\wbar{x}}
+\varepsilon(x-1)^{h}(-\wbar{x})^{\wbar{h}}\right)=\\
 =  \e^{i\theta_3(q)}w^{h}\wbar{w}^{\wbar{h}}
\left(\alpha+\beta(-x)^{h}(-\wbar{x})^{\wbar{h}}
+\gamma(x-1)^{h}(\wbar{x}-1)^{\wbar{h}}
\right. \\ \left.
+\rho(x-1)^{h}+\varepsilon(\wbar{x}-1)^{\wbar{h}}\right)\,.
\lab{eq:PPsiz}
\end{multline} 
Here we have used cyclic transformation laws  (\ref{eq:wxcyc}).

Comparing these two lines we obtain conditions: 
$\alpha=e^{-i\theta_3(q)}\beta$,
$\beta=e^{-i\theta_3(q)}\gamma$, $\gamma=e^{-i\theta_3(q)}\alpha$ and
$\rho=\varepsilon=0$. One can derive that $\exp(3i\theta_3(q))=1$
so $\theta_3(q)=\frac{2k\pi}{3}$ where $k=0,1,2$ ($k=0$ for physical
states). Thus the eigenstate of $\mathbb{P}$ \ci{Bartels:1999yt} 
reads
\begin{multline}
\Psi(\vec{z}_{10},\vec{z}_{20},\vec{z}_{30})=\\
=w^{h}\wbar{w}^{\wbar{h}}
\left(1+
\e^{i\frac{2\pi k}{3}}
(-x)^{h}(-\wbar{x})^{\wbar{h}}
+\e^{i\frac{4\pi k}{3}}
(x-1)^{h}(\wbar{x}-1)^{\wbar{h}}\right)\,,
\lab{eq:Psib}
\end{multline}
with $k=0,1,2$ and
where we have omitted the normalization constant.

Now we can act with a mirror permutation operator on (\ref{eq:Psym})
and test its eigenequation (\ref{eq:Msym}).
Using the mirror transformations (\ref{eq:wxmir}), similarly to 
(\ref{eq:PPsiz}),
we obtain the following relation
\begin{multline}
w^{h}\wbar{w}^{\wbar{h}}(-1)^{n_{h}}
\left(
\e^{i\frac{2\pi k}{3}}
+(-x)^{h}(-\wbar{x})^{\wbar{h}}
+\e^{i\frac{4\pi k}{3}}
(x-1)^{h}(\wbar{x}-1)^{\wbar{h}}\right)=\\
= \pm w^{h}\wbar{w}^{\wbar{h}}\left(1
+\e^{i\frac{2\pi k}{3}}
(-x)^{h}(-\wbar{x})^{\wbar{h}}
+\e^{i\frac{4\pi k}{3}}
(x-1)^{h}(\wbar{x}-1)^{\wbar{h}}\right)\,.
\lab{eq:MPsi}
\end{multline}
Comparing both sides
of (\ref{eq:MPsi})
gives  
$(-1)^{n_{h}}\exp(i\frac{2\pi k}{3})=\pm1$,
$(-1)^{n_{h}}=\pm\exp(i\frac{2\pi k}{3})$ and $(-1)^{n_{h}}=\pm1$
where the $\SL(2,\mathbb{C})$ Lorentz spins $n_h=h-\wbar h$.
These conditions are consistent with $k=0,\frac{3}{2}$. 
Only the first case
agrees with the cyclic permutation condition. As we can see for odd $n_{h}$
we have \emph{minus} sign, so taking into account colour factors $(-1)^{N}$,
solution (\ref{eq:MPsi}) is $C-$even. For even $n_{h}$ we have 
\emph{plus}
sign thus solution is $C-$odd. The last case is unnormalizable
because when $x\rightarrow0$ or $x\rightarrow1$ it does not vanish so
the norm, with $(s=0,\wbar s=0)$,  is divergent \ci{Bartels:1999yt}.

Using the duality symmetry \ci{Lipatov:1998as,Bartels:1999yt,Bartels:2001hw,
Vacca:2000bk,Kovchegov:2003dm},
which corresponds to  $h\rightarrow1-h$,
Bartels, Lipatov and Vacca constructed 
an eigenstate with $q_{3}=0$ and $C=-1$
\begin{multline}
\Psi(\vec{z}_{10},\vec{z}_{20},\vec{z}_{30}) =\\
=w^{h}\wbar{w}^{\wbar{h}}
x(1-x) \wbar{x}(1-\wbar{x})
\left(
\delta^{(2)} (x)-\delta^{(2)} (1-x)
+\frac{x^h \wbar{x}^h}{x^3 \wbar{x}^3}\delta^{(2)} \left(\frac{1}{x} \right)
\right)\,.
\lab{eq:BLV}
\end{multline} 
This wave-function cannot 
be constructed using the method described here because
it contains non-analytical function, the Dirac delta $\delta^{(2)}(x)$.

Now, let us take the second wave-function with five parameters \ie
for $q_{3}=0$ and $h\in\{0,1\}$. 
For $h=1$ 
the vector of linearly independent solutions  reads
\begin{equation}
u(x)=[1,(-x),(-x)\mbox{Log}(-x)+(x-1)\mbox{Log}(x-1)]
\end{equation}
 and for $\wbar{h}=0$ it is 
\begin{equation}
\wbar{u}(\wbar{x})=[\mbox{Log}(\wbar{x}-1),1,\mbox{Log}(-\wbar{x})]\,.
\end{equation}
Combining them we obtain
\begin{multline}
\Psi(\vec{z}_{10},\vec{z}_{20},\vec{z}_{30}) 
= w^{h}\wbar{w}^{\wbar{h}}\left(\alpha\mbox{Log}(\wbar{x}-1)+\beta
+\gamma(-x)\mbox{Log}(\wbar{x}-1)+\rho(-x)\right.\\
 \left.+\varepsilon((-x)\mbox{Log}(-\wbar{x})+(-x)\mbox{Log}(-x)
+(x-1)\mbox{Log}(x-1))\right)\,.
\lab{eq:ePsiq0}
\end{multline}
Like in the previous case we write the eigenequation for the 
cyclic permutation
\begin{multline}
 w^{h(=1)}\wbar{w}^{\wbar{h}(=0)}
\left((\alpha-\gamma-\varepsilon)\mbox{Log}(x-1)+(\rho-\beta)
+\alpha(-x)\mbox{Log}(\wbar{x}-1)\right.\\
  \left.+(-\beta)(-x)+\varepsilon(-x)\mbox{Log}(-x)
+\alpha(x-1)\mbox{Log}(-\wbar{x})\right.\\
  \left.+\varepsilon(-x)\mbox{Log}(-x)+\gamma\mbox{Log}(-\wbar{x})
+\varepsilon(x-1)\mbox{Log}(x-1)\right)=\\
 =  e^{i\theta_3(q)}w\left(\alpha\mbox{Log}(\wbar{x}-1)+\beta
+\gamma(-x)\mbox{Log}(\wbar{x}-1)+\rho(-x)\right.\\
  \left.+\varepsilon((-x)\mbox{Log}(-\wbar{x})+(-x)\mbox{Log}(-x)
+(x-1)\mbox{Log}(x-1))\right)\,.
\end{multline}
Thus we get conditions:
$\alpha=e^{-i\theta_3(q)}(\alpha-\gamma-\varepsilon)$,
$\beta=e^{-i\theta_3(q)}(\rho-\beta)$, $\gamma=e^{-i\theta_3(q)}\alpha$,
$\rho=e^{-i\theta_3(q)}(-\beta)$, $0=\e^{-i\theta_3(q)}(\gamma-\alpha)$
and $\varepsilon=e^{-i\theta_3(q)}\varepsilon$. 
We have two types of solutions.

The first one with $\theta_3(q)=0$ when  $\alpha=\gamma=-\varepsilon$
and $\beta=\rho=0$. It has a form
\begin{equation}
\Psi(\vec{z}_{10},\vec{z}_{20},\vec{z}_{30}) 
 =  
w\left((-x)\mbox{Log}((-\wbar{x})(-x))
+(x-1)\mbox{Log}((\wbar{x}-1)(x-1))\right)\,.
\lab{eq:Psiq0}
\end{equation}
We obtained in this way solutions with $\Log(x)-$terms which have
not been found before.The similar expressions 
were shown in \ci{DeVega:2001pu} 
as asymptotics of the $\oq{3}$ eigenfunction.
Acting with the mirror permutation operator on (\ref{eq:Psiq0}) we get
\begin{multline}
 \mathbb{M}w\left((-x)\mbox{Log}((-\wbar{x})(-x))
+(x-1)\mbox{Log}((\wbar{x}-1)(x-1))\right)=\\
 =  -w\left((-x)\mbox{Log}((-\wbar{x})(-x))
+(x-1)\mbox{Log}((\wbar{x}-1)(x-1))\right)\,.
\lab{eq:MPsilog}
\end{multline}
We obtained \emph{minus} sign so this state is also symmetric under
$C-$parity (\ie with $C=+1$).

Other solutions have $\theta_3(q)=2 \pi/3, 4\pi/3$.
Thus, $\alpha=\gamma=\varepsilon=0$ and 
$\rho=-\e^{-i\theta_3(q)}\beta$. The wave-function has a form
\begin{equation}
\Psi(\vec{z}_{10},\vec{z}_{20},\vec{z}_{30}) =
w\, (1+x\,\e^{i \theta_3(q)})\,.
\lab{eq:Psiq0b}
\end{equation}
These solutions are not eigenstates of the $\mathbb{M}$ operator.

\section{Known features of the spectrum}

In this Section we describe the spectra of the conformal charges
obtained by numerical solutions \ci{Derkachov:2002wz,Kotanski:2001iq}.
To this end we apply the Baxter operator method and the Janik-Wosiek
method described in the previous Section.
For $N=3$ reggeized gluons the Hamiltonian method 
and the Baxter operator method give the same results.

\subsection{Advantages of $Q-$Baxter operator method}

In order to find the spectrum of the conformal charges
for higher $N$ and the Reggeon energy $E_N$
one can use the $Q-$Baxter operator method \ci{Baxter,Derkachov:2002wz}. 
The Baxter operator $\mathbb{Q}(u,\wbar{u})$
depends on two complex spectral parameters $u$, $\wbar u$
and commutes with the integrals  of motion, $q_k$ and $\wbar q_k$.
It satisfies the Baxter equation:
\begin{equation}
\ot{N}(u) \mathbb{Q}(u,\wbar{u})  =
(u + i s)^N \mathbb{Q}(u+i,\wbar{u})  +
(u - i s)^N \mathbb{Q}(u-i,\wbar{u})  \,,
\lab{eq:Baxeq}
\end{equation}
\begin{equation}
\otb{N}(\wbar{u}) \mathbb{Q}(u,\wbar{u})  =
(\wbar{u} + i \wbar{s})^N \mathbb{Q}(u,\wbar{u}+i)  +
(\wbar{u} - i \wbar{s})^N \mathbb{Q}(u,\wbar{u}-i)  \,,
\lab{eq:Baxbeq}
\end{equation}
where $\ot{N}(u)$ is the auxiliary transfer matrix
\begin{equation}
\ot{N}(u)=2u^{N}+\oq{2} u^{N-2}+ \ldots +\oq{N}
\lab{eq:tnu}
\end{equation}
and similarly $\otb{N}(\wbar{u})$.
We solve the Baxter equations \ci{Derkachov:2002wz} using the integral ansatz
which changes Equations \ref{eq:Baxeq} and \ref{eq:Baxbeq} to differential equations.
The latter ones can be solved analogically to the Janik-Wosiek method
which gives $Q-$Baxter quantization conditions.

The Reggeon energy is determined by eigenvalues of the Baxter operator,
$Q_{q_k,\wbar q_k}(u,\wbar u)$ and can be written as
\begin{multline}
E_N(q,\bar q) =- \Im\frac{d}{du}\ln \bigg[u^{2N}Q_{q,\bar
q}(u+i(1-s),u+i(1-\bar s))\,\\
\times Q_{-q,-\bar q}(u+i(1-s),u+i(1-\bar s))\bigg]\bigg|_{u=0} \,,
\lab{eq:enQ2}
\end{multline}
with
\begin{equation}
\pm q=(q_2,\pm q_3,\ldots,(\pm)^N q_N) 
\lab{eq:pmq}
\end{equation}
while the quasimomentum $\theta_N$
\begin{equation}
  \theta_N
= i \ln \frac{ Q_{q,\bar q}(is,i\wbar{s})}{
Q_{q,\wbar q}(-is,-i\wbar{s})}\,.
\lab{eq:quasQ}
\end{equation}

\subsection{Trajectories}
Solving the $Q-$Baxter quantization conditions 
\ci{Derkachov:2002wz}
we obtain continuous trajectories
in the space of conformal charges.
They are built of points, $(q_2(\nu_h),\ldots,q_N(\nu_h))$ which
 depend on a continuous 
real parameter  $\nu_h$ entering $q_2$, (\ref{eq:q2}) and 
(\ref{eq:hpar}).
In order to label the trajectories 
we introduce the set of the integers 
\begin{equation}
\mybf{\ell}=\{\ell_1,\ell_2,\ldots,\ell_{2(N-2)}\}
\lab{eq:dell}
\end{equation}
which parameterize one specified point on each trajectory 
for given $h$.
Specific examples in the following sections will further clarify
this point. 

Next we calculate the observables
along these trajectories, 
namely the energy (\ref{eq:Schr}) 
and the quasimomentum (\ref{eq:quask}). 
The quasimomentum is constant for all points 
situated on a given trajectory.
The minimum of the energy, which means the maximal intercept,
for almost all trajectories is located
at $\nu_h=0$.
It turns out that the energy behaves around $\nu_h=0$ like
\begin{equation}
E_N(\nu_h;\mybf{\ell}^{\rm ground})
=E_N^{\rm ground}+\sigma_N {\nu_h}^2+{\cal O}({\nu_h}^2)
\lab{eq:Enu}
\end{equation}
Thus, the ground state along its trajectory is gapless 
and the leading contribution
to the scattering amplitude around $\nu_h$ may be rewritten as
a series in the strong coupling constant:
\begin{equation}
{\cal A}(s,t) \sim -i s 
\sum _{N=2}^\infty (i \asbar)^N
\frac{s^{-\asbar E_N^{\rm ground}/4}}{(\asbar\sigma_N
\ln s)^{1/2}}\,\xi_{A,N}(t)
\xi_{B,N}(t)\,,
\lab{eq:amp}
\end{equation}
where  $\asbar=\alpha_s N_c/\pi$ 
and $\xi_{X,N}(t)$ are the impact factors
corresponding to the overlap between the wave-functions of scattered
particle with the wave-function of $N-$Reggeons, whereas 
$\sigma_N$ measures the dispersion of the energy on the
the trajectory around $\nu_h=0$.

On the other hand,
the energy along the trajectories 
grows with $\nu_h$ and for $|\nu_h| \rightarrow \infty$ 
and finally, we have 
$E_N(\nu_h;{\mybf \ell}) \sim \ln {\nu_h}^2$. 
These parts of the trajectory give the lowest contribution to the
scattering amplitude.

\subsection{Symmetries}
The spectrum of quantized charges $q_2,\ldots,q_N$ is degenerate.
This degeneration is caused by two symmetries:
\begin{equation}
q_k \leftrightarrow (-1)^k q_k
\lab{eq:qkmsym}
\end{equation}
which comes from
invariance of the Hamiltonian under mirror permutations of particles, 
(\ref{eq:Msym}), and
\begin{equation}
q_k \leftrightarrow  \wbar q_k
\lab{eq:qkcsym}
\end{equation}
which is 
connected with
the symmetry under interchange of the $z-$ and 
$\wbar z-$sectors.
Therefore, the four points, \ie. $\{q_k\}$, $\{(-1)^k q_k\}$,
$\{{q_k}^\ast\}$ and $\{(-1)^k {q_k}^{\ast}\}$ with $k=2,\ldots,N$, 
are related and all of them satisfy the quantization conditions 
(\ref{eq:azao}), (\ref{eq:aoai})
and have the same energy.

\subsection{Descendent states}

Let us first discuss
the spectrum along the trajectories
with the highest conformal charge $q_N$
equal zero
for arbitrary $\nu_h \in \mathbb{R}$.
It turns out \ci{Vacca:2000bk,Lipatov:1998as,Bartels:1999yt,Derkachov:2002wz} 
that the wave-functions of these states
are built of $(N-1)-$particle states. Moreover, their energies
\ci{Korchemsky:1994um} are also equal to the energy of the ancestor
$(N-1)-$particle states: 
\begin{equation}
E_N(q_2,q_3,\ldots,q_N=0)=E_{N-1}(q_2,q_3,\ldots,q_{N-1})\,.
\lab{eq:Edes}
\end{equation}
Thus, we call them the descendent states
of the $(N-1)-$particle states. 

Generally, for odd $N$, the descendent state $\Psi_N^{(q_N=0)}$
with the minimal energy $E_N(q_N=0)=0$ 
has 
for $q_2=0$, (\ie for $h=0,1$),  the remaining 
integrals of motion 
$q_3=\ldots =q_N=0$ as well.
For $h=1+i \nu_h$, \ie $q_2 \ne 0$,
the odd conformal charges $q_{2k+1}=0$ with $k=1,\ldots,(N-1)/2$
while the even ones $q_{2k}\ne 0$ and depend on $\nu_h$.

On the other hand, for even $N$, the eigenstate with the minimal energy 
$\Psi_N^{(q_N=0)}$  is the descendent state of the $(N-1)-$particle 
state 
which has minimal energy with $q_{N-1} \ne 0$.  Thus, 
${E_N^{min}(q_N=0)=E_{N-1}^{min}(q_{N-1}\ne0)>0}$.

Studying more thoroughly this problem one can obtain
\ci{Vacca:2000bk,Derkachov:2002wz}
a relation between the quasimomentum $\theta_N$ of the descendent state 
and the ancestor one  $\theta_{N-1}$, which takes the following form
\begin{equation}
\e^{i \theta_N} \bigg|_{q_{N}=0}=
-\e^{i \theta_{N-1}}=
(-1)^{N+1}\,.
\lab{eq:quasdes}
\end{equation} 

Additionally, one can define 
a linear operator $\Delta$  
\ci{Vacca:2000bk,Derkachov:2002wz}
that maps 
the subspace $V_{N-1}^{(q_N-1)}$ of the $(N-1)-$particle 
ancestor eigenstates 
with the quasimomentum $\theta_{N-1}=\pi N$
into the $N-$particle descendent states with $q_N=0$
and $\theta_N=\pi (N+1)$ as
\begin{equation}
\Delta:
\quad
V_{N-1}^{(\theta_{N-1}=\pi N)} \to 
V_{N}^{(\theta_N=\pi (N+1))} \;.
\lab{eq:Delta}
\end{equation}
It turns out that this operator is nilpotent 
for the eigenstates 
which form trajectories
\ci{Vacca:2000bk}, \ie $\Delta^2 \Psi=0$.
Thus, the descendent-state trajectory 
can not be ancestor trajectory for $(N+1)-$particle states.
However, it is possible to built a single  state \ci{Derkachov:2002wz}
with $q_2=q_3=\ldots=q_N=0$, \ie for only one point $\nu_h=0$,
that has $E_N=0$ and the eigenvalue of Baxter $Q-$operator
defined as
\begin{equation}
Q_N^{q=0}(u,\wbar u) \sim \frac{u-\wbar u}{\wbar u^N}\,,
\lab{eq:Qq0}
\end{equation}
where a normalization factor was omitted.
For $N=3$ this state corresponds to the wave-function defined in 
(\ref{eq:Psiq0}).

Additional 
examples of the descendent states for $N=3$ will be 
described later in the next sections.

\section{Numerical results}

\subsection{Quantum numbers of the $N=3$ states}

In this section we present the spectrum of $q_3$  
for three reggeized gluons. 
For the first time such solutions  for $\Re[q_3]=0$
were obtained 
in \ci{Wosiek:1996bf}. 
The authors of Ref. \ci{Wosiek:1996bf}
used the method of the $\oq{3}$ eigenfunctions
described in Section 5. 
Similar quantization condition was also constructed in
\ci{Lipatov:1998as}.
Solutions with
$\Re[q_3] \ne 0$ were found in \ci{Praszalowicz:1998pz,Kotanski:2001iq}.
Moreover, in the latter paper 
the solutions with $n_h \ne 0$ are described.

The first solution
using the Baxter $Q-$operator method
was described
in Ref.
\ci{Derkachov:2002wz}.
Later,  similar results were also obtained in
Ref. \ci{deVega:2002im}.
It turns out that the Baxter $Q-$operator method \ci{Derkachov:2002wz}
for $N=3$
is equivalent to the method of the $\oq{3}$ eigenfunctions
described in Section 5.
For higher $N>3$ only the Baxter $Q-$operator method was used to find
the quantization values of $q_N$ and the energy.

\subsection{Lattice structure}

Solving the quantization conditions (\ref{eq:azao}) and  (\ref{eq:aoai})
for $N=3$
and for $h=\frac{1+n_h}{2}$, \ie with $\nu_h=0$,
we reconstruct the full spectrum of $q_3$.
It is convenient to show the spectrum in terms of $q_3^{1/3}$ rather 
than $q_3$. 
Since $q_3^{1/3}$  is a multi-valued function of complex $q_3$,
each eigenstate is represented on the complex
$q_3^{1/3}-$plane by $N=3$ different points.
Thus, the spectrum is symmetric under the
transformation
\begin{equation}
q_3^{1/3} \leftrightarrow \exp(2 \pi i k/3) q_3^{1/3},
\quad
\mbox{where}
\quad
0<k<3\,.
\end{equation} 
Additionally, mirror symmetry (\ref{eq:qkmsym}) results in
a more regular structure.
For the total  $\SL(2,\mathbb{C})$ spin of the system 
$h=1/2$, which means $n_h=0$ and $\nu_h=0$,
we present the spectrum in Fig.~\ref{fig:N3q0}.
\begin{figure}[ht]
\centerline{
\begin{picture}(150,65)
\put(15,2){\epsfysize6.0cm \epsfbox{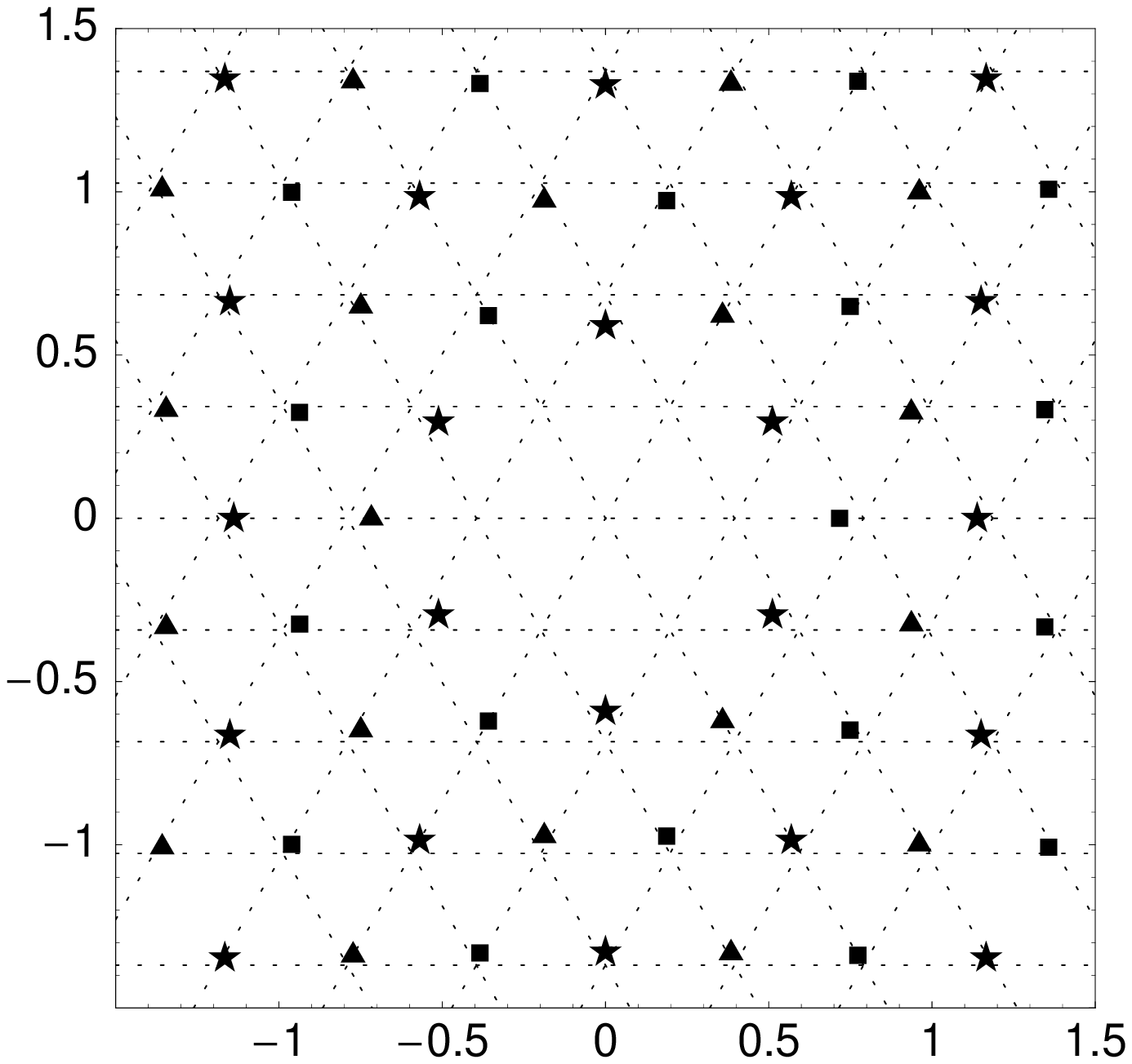}}
\put(78,2){\epsfysize6.0cm \epsfbox{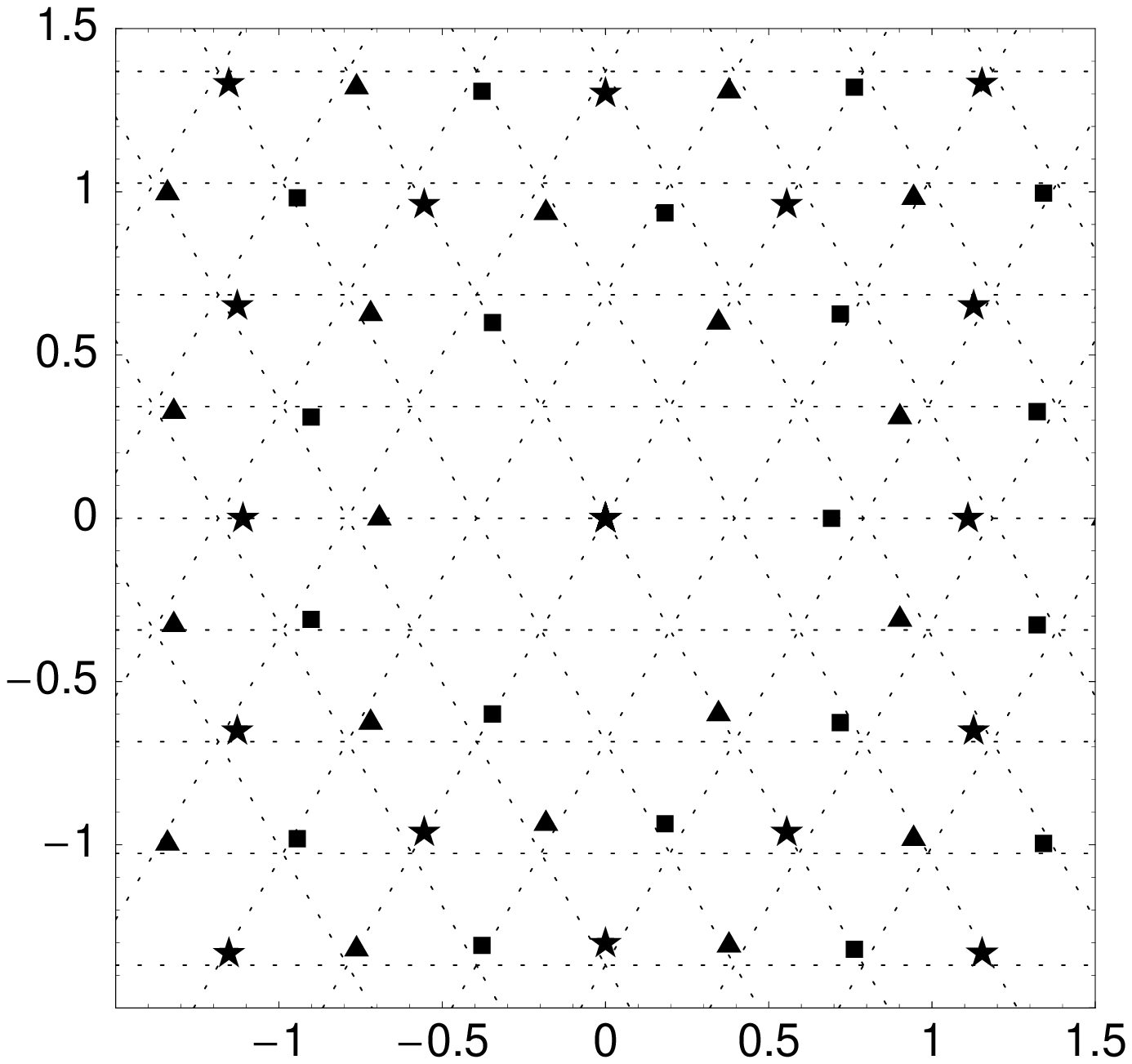}}
\put(42,0){$\Re[q_3^{1/3}]$}
\put(105,0){$\Re[q_3^{1/3}]$}
\put(13,27){\rotatebox{90}{$\Im[q_3^{1/3}]$}}
\put(76,27){\rotatebox{90}{$\Im[q_3^{1/3}]$}}
\end{picture}
}
\caption[The spectrum of $q_3^{1/3}$ for the 
system of $N=3$ particles for $h=\frac{1}{2}$ and $h=1$]
{The spectrum of quantized $q_3^{1/3}$ for the 
system of $N=3$ particles.
On the left 
the total $\SL(2,\mathbb{C})$ spin of the system is equal to $h=\frac{1}{2}$,
while on the right $h=1$.
Different symbols stand for different quasimomenta $\theta_3$:
{\it stars} $\theta_3=0$
{\it boxes} $\theta_3=4 \pi/3$
{\it triangles} $\theta_3=2 \pi/3$.}
\lab{fig:N3q0}
\lab{fig:N3q1}
\end{figure}
One can easily notice that the spectrum has the structure
close to the equilateral triangle lattice
of the leading order WKB approximation (\ref{eq:korq3}). 
Indeed,
apart from a few points close to the origin, the quantized values of 
$q_3^{1/3}$ are located almost exactly at the vertices of the WKB
lattice. 
The WKB formula \ci{Derkachov:2002pb} gives
\begin{equation}
\left[q_{3}^{\rm WKB}(\ell_1,\ell_2)\right]^{1/3}
=\Delta_{N=3}
\left(\frac12\ell_1+i\frac{\sqrt{3}}2\ell_2\right),
\lab{eq:q3-WKB}
\end{equation}
where $\ell_1$ and $\ell_2$ are integers, 
such that their sum $\ell_1+\ell_2$ is
even. 
Here the lattice spacing is denoted by 
\begin{equation}
\Delta_{3}
=\left[\frac{3}{4^{1/3}\pi}
\int_{-\infty}^1\frac{dx}{\sqrt{1-x^3}}\right]^{-1}
=\frac{\Gamma^3(2/3)}{2\pi}=0.395175\ldots \,.
\end{equation}
The lattice of
$q_3^{1/3}$ 
extends onto the whole complex plane except the interior of the
disk with the radius $\Delta_{3}$:
\begin{equation}
|q_3^{1/3}| > \Delta_{3}\,
\end{equation}
situated at $q_3=0$.

\begin{figure}[ht]
\centerline{
\begin{picture}(150,65)
\put(15,2){\epsfysize6.0cm \epsfbox{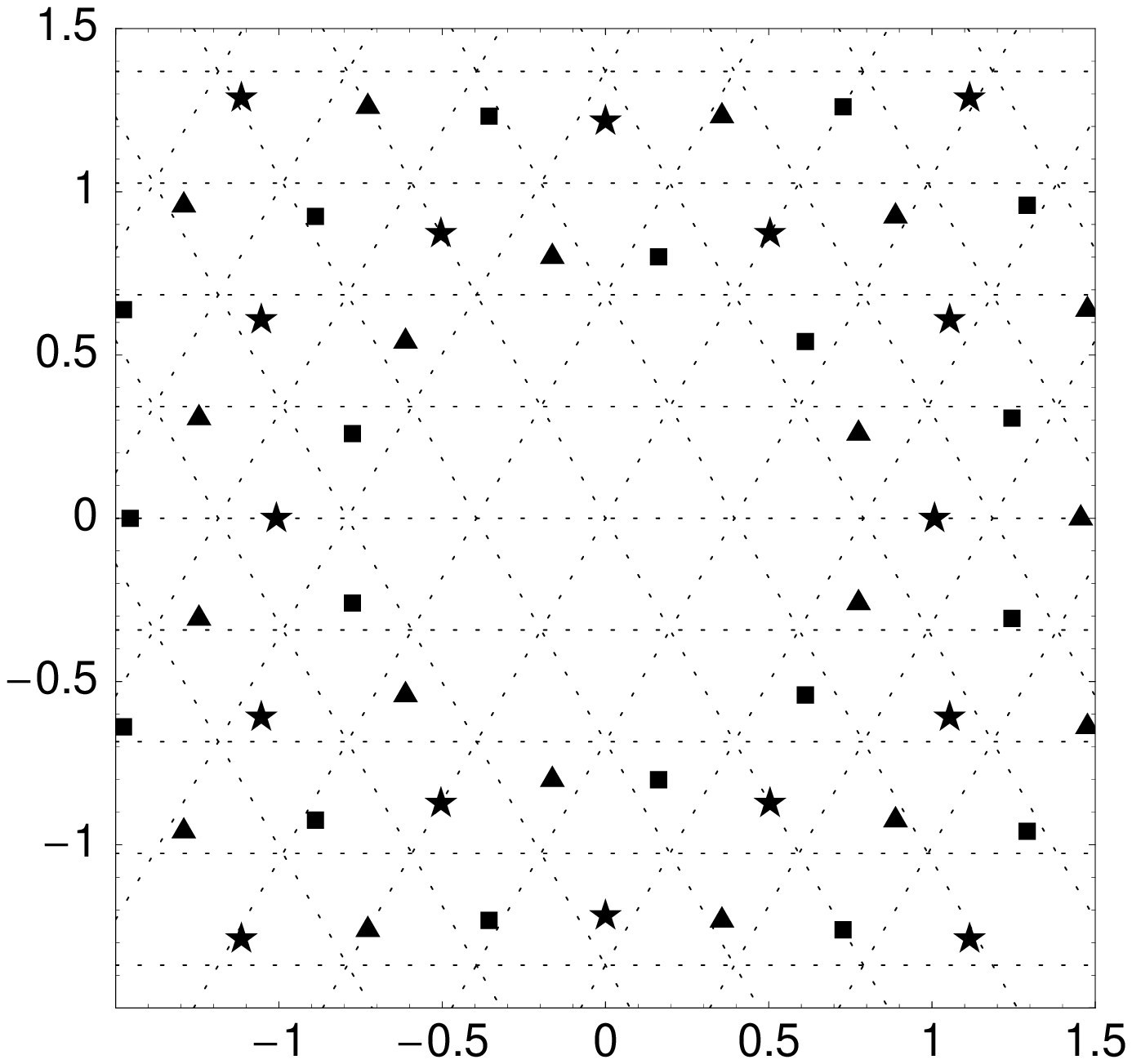}}
\put(78,2){\epsfysize6.0cm \epsfbox{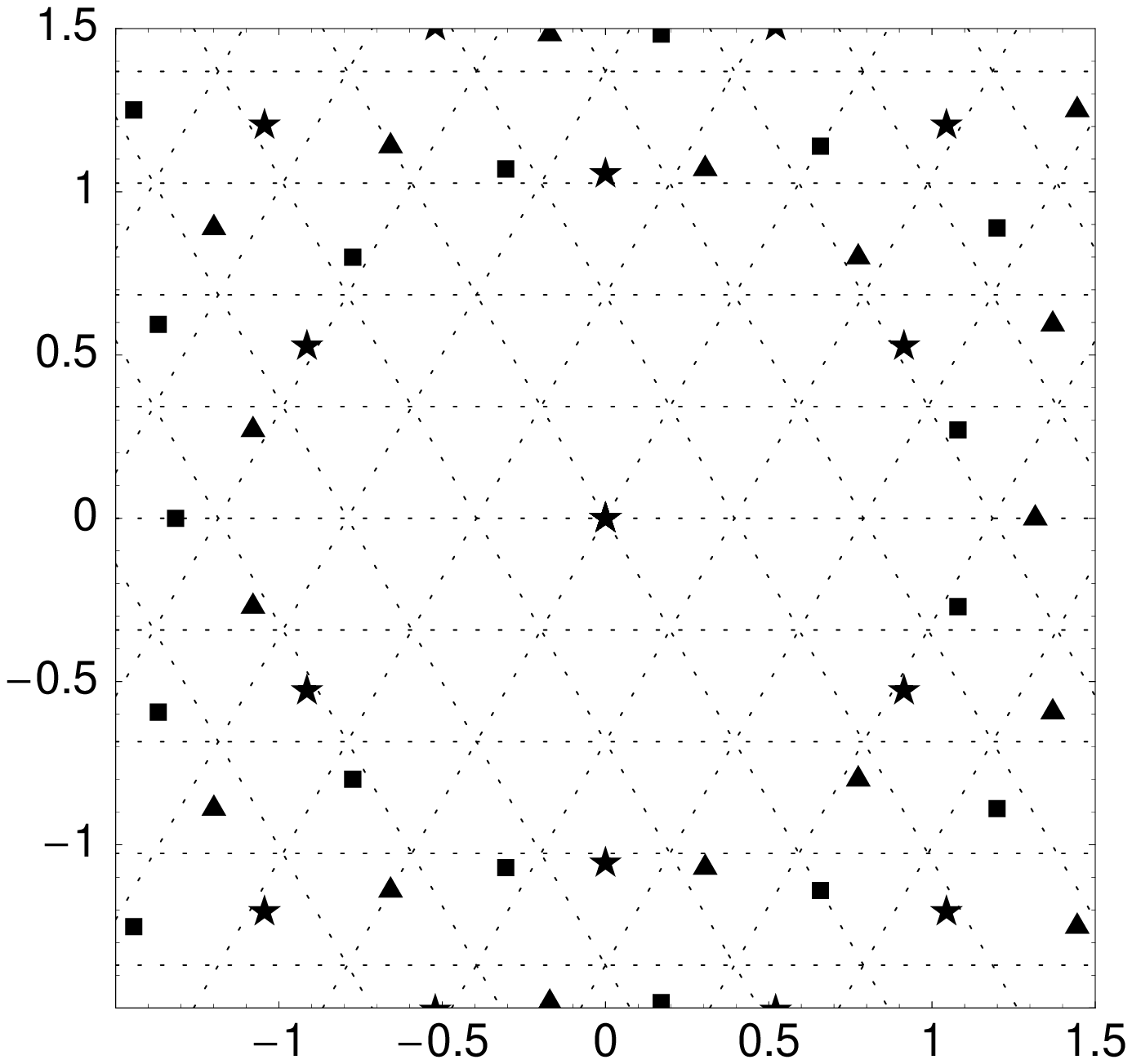}}
\put(42,0){$\Re[q_3^{1/3}]$}
\put(105,0){$\Re[q_3^{1/3}]$}
\put(13,27){\rotatebox{90}{$\Im[q_3^{1/3}]$}}
\put(76,27){\rotatebox{90}{$\Im[q_3^{1/3}]$}}
\end{picture}
}
\caption[The spectrum of $q_3^{1/3}$ for the 
system of $N=3$ particles for $h=\frac{3}{2}$ and $h=2$]
{The spectrum of quantized $q_3^{1/3}$ for the 
system of $N=3$ particles.
On the left 
the total $\SL(2,\mathbb{C})$ spin of the system is equal to $h=\frac{3}{2}$,
while on the right $h=2$.
Different symbols stand for different quasimomenta $\theta_3$:
{\it stars} $\theta_3=0$
{\it boxes} $\theta_3=4 \pi/3$
{\it triangles} $\theta_3=2 \pi/3$.}
\lab{fig:N3q2}
\lab{fig:N3q3}
\end{figure}

In accordance with (\ref{eq:q3-WKB}),
a pair of integers $\ell_1$ and $\ell_2$
parameterize 
the quantized values of $q_3^{1/3}$. Going further,
one can calculate the quasimomentum as a function of 
$\ell_1$ and $\ell_2$. It has a following form
\begin{equation}
\theta_3(\ell_1,\ell_2)=\frac{2\pi}3\ell_1\quad ({\rm mod}~2\pi)
\,.
\lab{eq:quasi-3}
\end{equation}
Thus, as we can see states with the same value of 
$\Re[q_3^{1/3}]$ have the same quasimomentum.
In Fig.~\ref{fig:N3q1}, different quasimomenta are
distinguished by {\it stars}, {\it boxes} and {\it triangles}.

The same lattice structure is exhibited by the spectra
with different $n_h$. 
However, they have different corrections to the 
leading order WKB approximation 
for $q_3^{1/3}$. These spectra are presented in 
Figs.\ref{fig:N3q1}-\ref{fig:N3q3}. 
The corrections to the lattice structure 
depend on $q_2$ as seen in (\ref{eq:korq3}).
Since the WKB lattice is 
obtained 
in the leading order of 
the expansion
for large conformal charges, $1 \ll |q_2^{1/2}| \ll |q_3^{1/3}|$, 
the corrections are bigger for lower $|q_3^{1/3}|$.
Later, we shall discuss some other features of the corrections
to the WKB leading order approximation.

As we can see in Figs.\ \ref{fig:N3q1}-\ref{fig:N3q3} 
for $h \in \mathbb{Z}$ we have additionally
trajectories with  $q_3=0$. They are called
the descendent states because 
their spectra are related to the spectra for the $N-1=2$ Reggeon states. 
We discuss this point further below.

\subsection{Trajectories in $\nu_h$}

In the previous Section we considered the dependence of $q_3$ 
on $n_h$ for $\nu_h=0$.
However, the spectrum of conformal charges also depends 
on the continuous parameter $\nu_h$ with
$h=\frac{1+n_h}{2}+i \nu_h$. 
It turns out that 
the spectrum is built of trajectories 
parameterized by real parameter $\nu_h$.
Each trajectory
crosses one point ({\it star}, {\it box}, {\it triangle})
in Figs.\ \ref{fig:N3q1}-\ref{fig:N3q3}. 
An example of three such trajectories is presented in Fig.~\ref{fig:traj-3D}.
They are numbered by $(\ell_1,\ell_2)=(0,2)$, $(2,2)$ and $(4,2)$ 
whereas
they quasimomentum $\theta_3(\ell_1,\ell_2)=0$, $4 \pi/3$ and $2 \pi/2$, 
respectively.

\begin{figure}[ht]
\vspace*{3mm}
\centerline{
\psfrag{nu_h}[cc][bc]{$\mbox{\large$\nu_h$}$}
\psfrag{Im(q3^(1/3))}[cc][cc]{$\Im[q_3^{1/3}]$}
\psfrag{Re(q3^(1/3))}[cc][bc]{$\Re[q_3^{1/3}]$}
{\epsfysize6.5cm \epsfbox{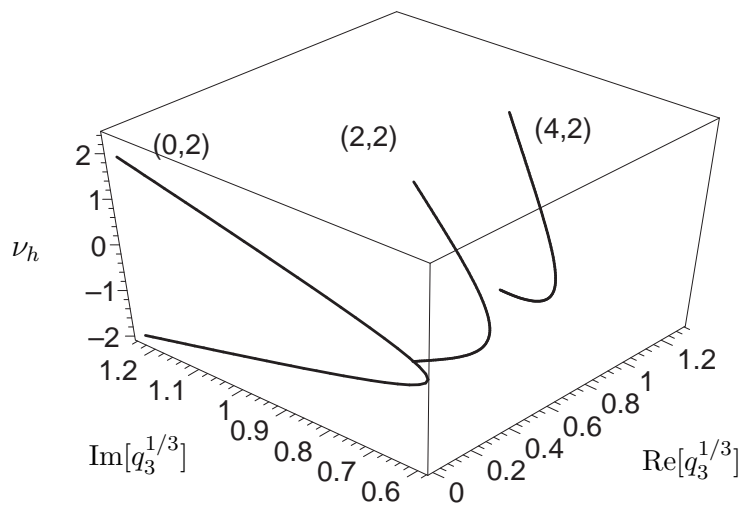}}}
\caption[The dependence of quantized $q_3(\nu_h;\ell_1,\ell_2)$ 
on the total spin $h=1/2+i\nu_h$.]{The dependence of 
quantized $q_3(\nu_h;\ell_1,\ell_2)$ on the total spin $h=1/2+i\nu_h$.
Three curves correspond to the trajectories 
with $(\ell_1,\ell_2)=(0,2)\,, (2,2)$
and $(4,2)$.}
\lab{fig:traj-3D}
\end{figure}

\begin{figure}[ht]
\centerline{
\begin{picture}(150,65)
\put(15,2){\epsfysize6.0cm \epsfbox{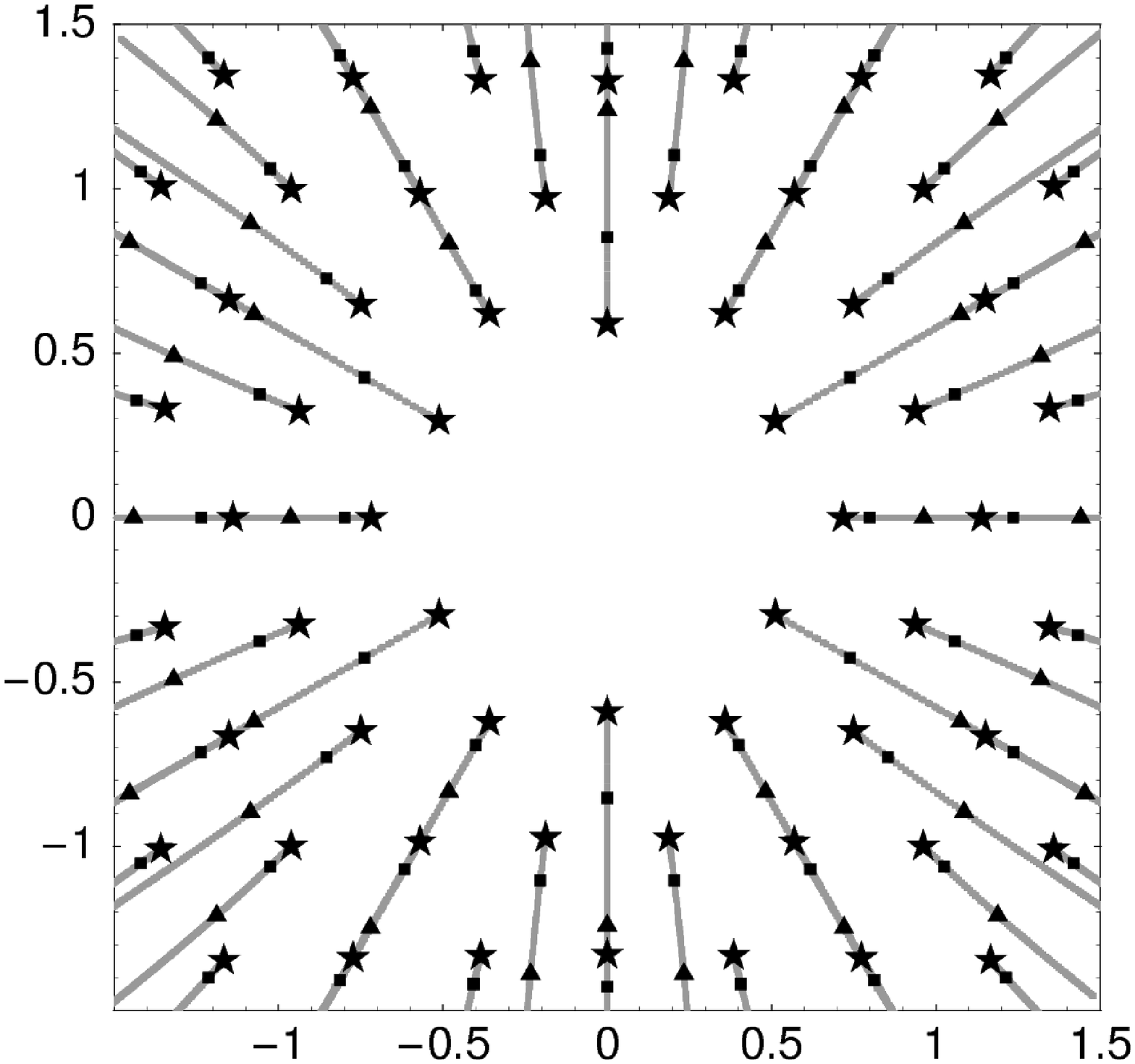}}
\put(78,2){\epsfysize6.0cm \epsfbox{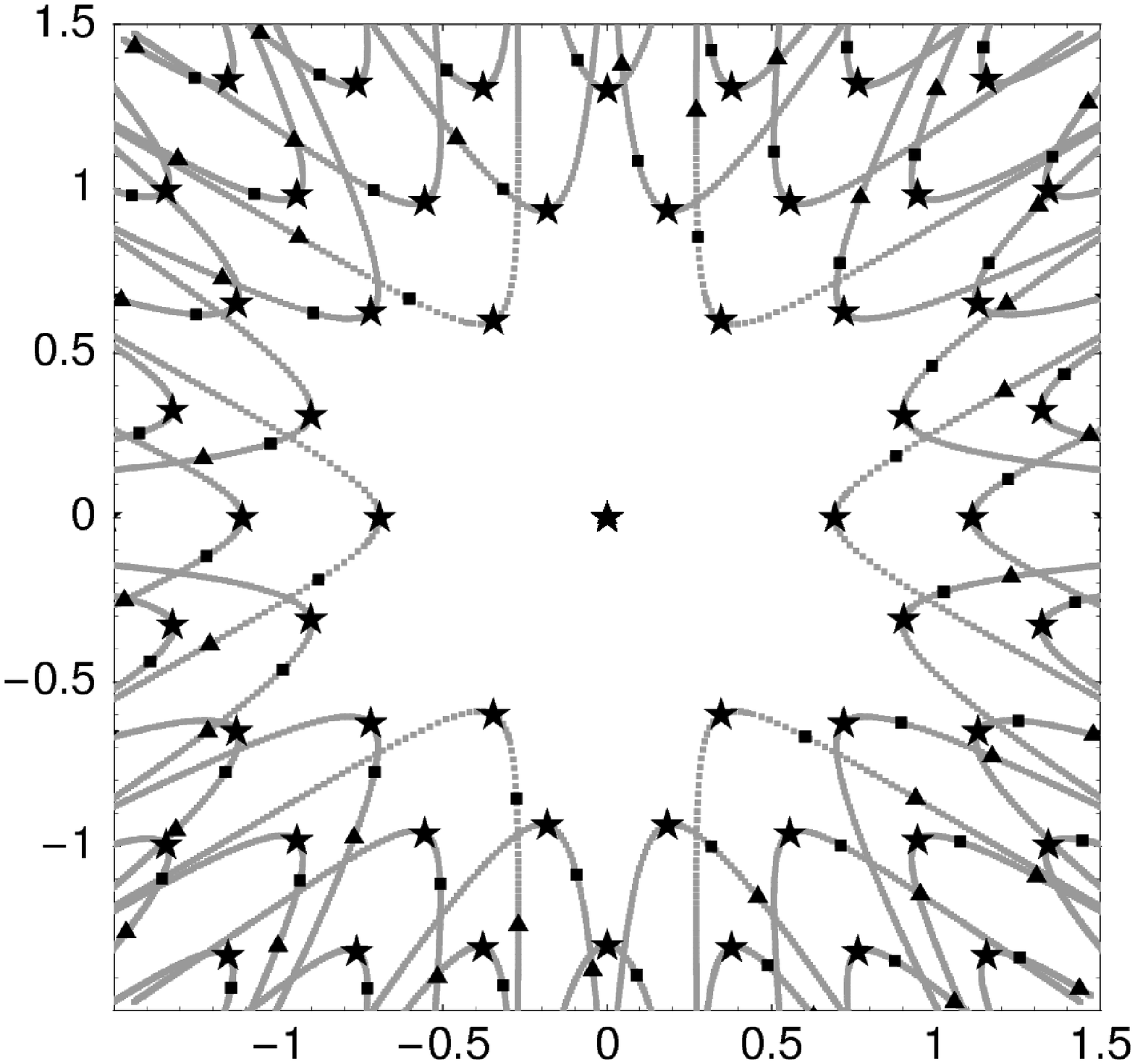}}
\put(42,0){$\Re[q_3^{1/3}]$}
\put(105,0){$\Re[q_3^{1/3}]$}
\put(13,27){\rotatebox{90}{$\Im[q_3^{1/3}]$}}
\put(76,27){\rotatebox{90}{$\Im[q_3^{1/3}]$}}
\end{picture}
}
\caption[The trajectories of $q_3^{1/3}$ projected on $\nu_h=0$
for $n_h=0$ and $n_h=1$.]
{The trajectories of $q_3^{1/3}$ projected on $\nu_h=0$.
On the left panel $h=\frac{1}{2}+ i \nu_h$, while on 
the right one $h=1+ i \nu_h$. {\it stars} denotes $\nu_h=0$,
{\it boxes} $\nu_h=1$ and {\it triangles} $\nu_h=2$.}
\lab{fig:prtraj}
\end{figure}

The trajectories cumulate at $\nu_h=0$. When we increase $\nu_h$, 
$q_3^{1/3}$ tends to infinity and the structure
of quantized charges 
starts to be less regular, especially for trajectories with lower 
$|q_3^{1/3}|$. 
We can see this in Fig.~\ref{fig:prtraj} where we project trajectories
with $h=\frac{1}{2}+i \nu_h$
on the $\nu_h=0$ plane. Here {\it stars} denote point with $\nu_h=0$,
{\it boxes} with $\nu_h=1$ and {\it circles} $\nu_h=2$.
Grey lines are drawn to show the projection of the trajectories for 
intermediate values of $\nu_h$.

\begin{figure}[ht]
\centerline{
\begin{picture}(150,65)
\put(15,2){\epsfysize6.0cm \epsfbox{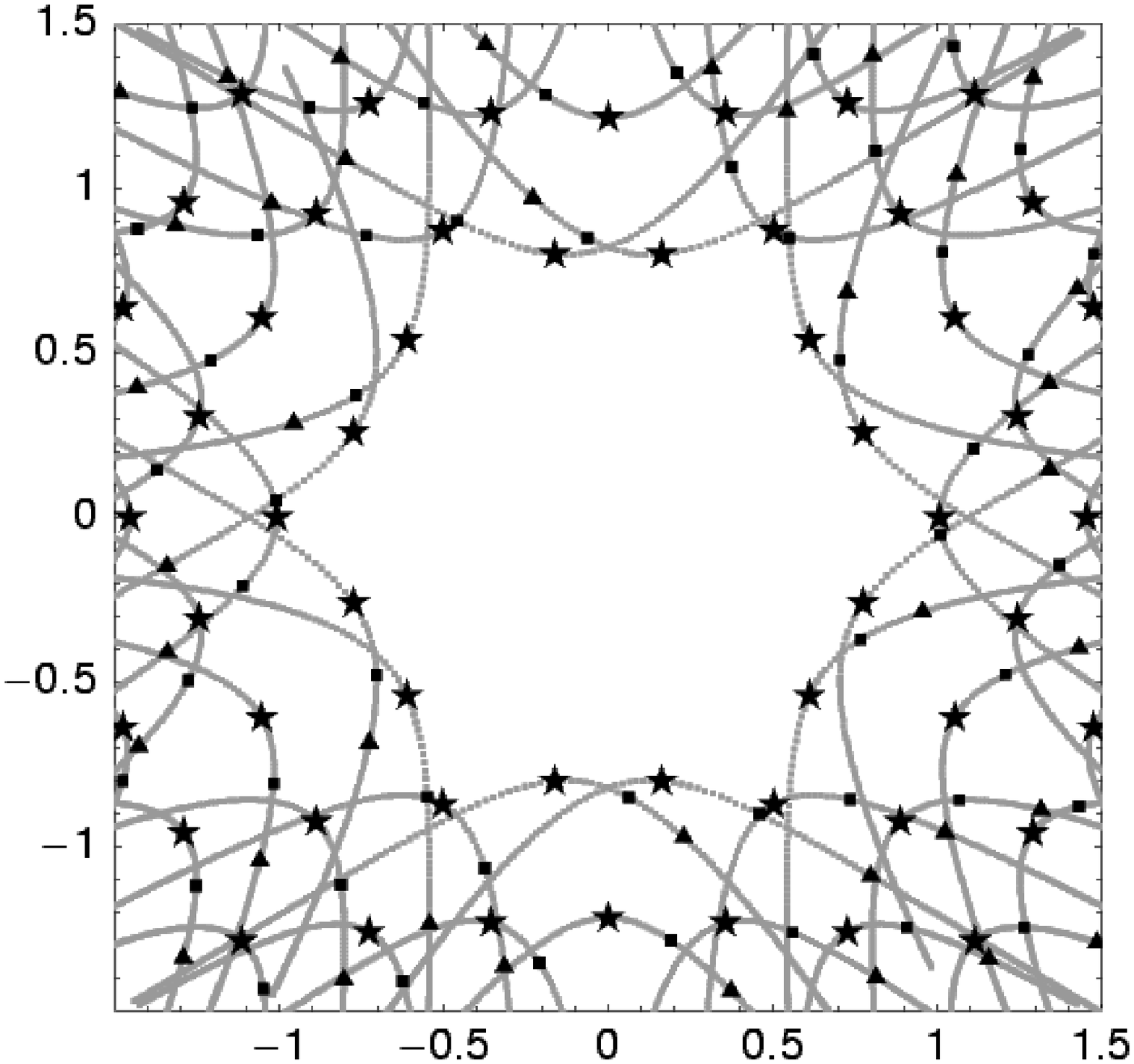}}
\put(78,2){\epsfysize6.0cm \epsfbox{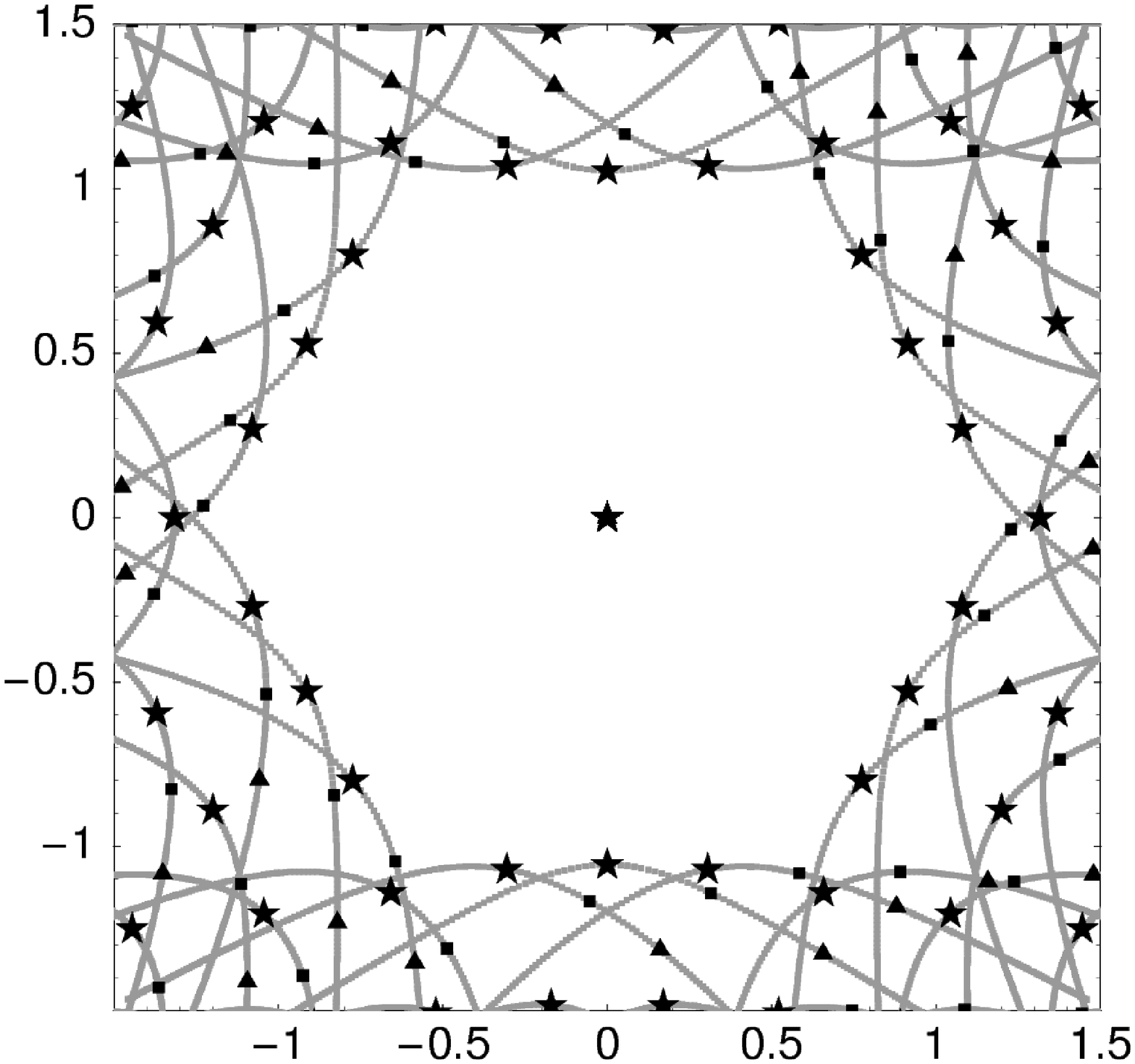}}
\put(42,0){$\Re[q_3^{1/3}]$}
\put(105,0){$\Re[q_3^{1/3}]$}
\put(13,27){\rotatebox{90}{$\Im[q_3^{1/3}]$}}
\put(76,27){\rotatebox{90}{$\Im[q_3^{1/3}]$}}
\end{picture}
}
\caption[The trajectories of $q_3^{1/3}$ projected on $\nu_h=0$
for $n_h=2$ and $n_h=3$.]
{The trajectories of $q_3^{1/3}$ projected on $\nu_h=0$.
On the left panel $h=\frac{3}{2}+ i \nu_h$, while on 
the right one $h=2+ i \nu_h$. {\it stars} denotes $\nu_h=0$,
{\it boxes} $\nu_h=1$ and {\it triangles} $\nu_h=2$.}
\lab{fig:prtraj2}
\end{figure}
For $n_h \ne 0$ we notice that the spectra start to rotate with $\nu_h$.
In Fig.~\ref{fig:prtraj} and \ref{fig:prtraj2}  
we present trajectories with positive $n_h=0,1,2,3$.
Due to the symmetry (\ref{eq:qkcsym}), which means 
$h\rightarrow 1-h^{\ast}$ or 
$n_h \rightarrow - n_h$ with $q_3 \rightarrow q_3^{\ast}$,
or equivalently,   
$\nu_h \rightarrow -\nu_h$ but $q_3 \rightarrow q_3$,
the spectrum for the negative $n_h$  is the
same as for the positive ones but
it rotates in the opposite direction with $\nu_h$.

Some of the results presented in this Section were found in earlier works
\ci{Korchemsky:1999is,Kotanski:2001iq}.
Trajectories with 
quasimomentum $\theta_3=0$ and
$n_h=0$ were obtained in Refs.~\ci{Korchemsky:1999is,Kotanski:2001iq}.
The case for $h=2+i \nu_h$ with 
quasimomentum $\theta_3=0$ was discussed in Ref. \ci{Kotanski:2001iq}.
In this work we additionally analyse the spectra for 
$h=1+i\nu_h$ and 
$h=3/2+i\nu_h$.

\subsection{Energy and dispersion}

For all trajectories 
in $(q_2,q_3)-$space 
we can calculate the energy of the reggeized gluons
using Eq. (\ref{eq:enQ2}). 
Example of the energy spectrum for trajectories
from Fig.~\ref{fig:traj-3D} with $h=\frac{1}{2}+i \nu_h$ is shown in
Fig.~\ref{fig:energy3}.
\begin{figure}[ht]
\vspace*{3mm}
\centerline{
\psfrag{-E_3/4}[cc][cc]{$-E_3/4$}
\psfrag{nu_h}[cc][bc]{$\mbox{\large$\nu_h$}$}
{\epsfysize7cm \epsfbox{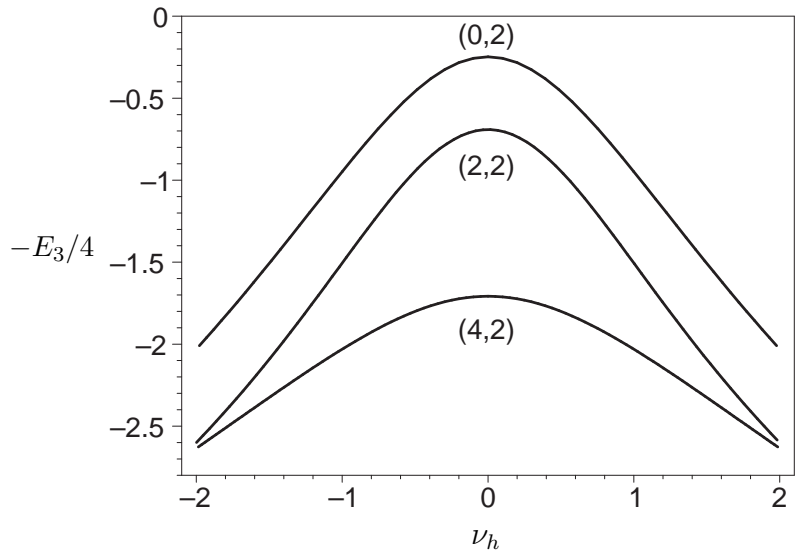}}}
\caption{The energy spectrum corresponding to three trajectories shown
in Fig.~\ref{fig:traj-3D}.
The ground state is located on the $(0,2)-$trajectory at $\nu_h=0$.}
\lab{fig:energy3}
\end{figure}

The energy along the trajectories is a continuous gapless 
function of $\nu_h$.
As we can see the energy $E_3$ grows with rising $|\nu_h|$.
For $n_h=0$ it has a minimum value 
$\mbox{min}_{\nu_h} E_3(\nu_h;\ell_1,\ell_2)$ at $\nu_h=0$.
In the case $n_h \ne 0$, due the bending of the trajectories 
some minima of the energy are moved away from $\nu_h=0$ 
\ci{Kotanski:2001iq}. 
However, the ground state corresponds to the point(s) on the plane of 
$q_3^{1/3}$ (see Fig.~\ref{fig:N3q0}) closest to the origin.
For $N=3$ the ground state is located on the $(0,2)-$trajectory at 
$\nu_h=0$ and $n_h=0$  with quasimomentum equal $\theta_3=0$.
Due to the symmetry (\ref{eq:qkmsym})
it is doubly-degenerated and its conformal charge and energy 
take the following values:
\begin{equation}
iq_3^{\rm ground}=\pm 0.20526\ldots\,,\qquad E_3^{\rm ground}=0.98868\ldots\,.
\lab{eq:N3-ground}
\end{equation}
In the vicinity of $\nu_h=0$ 
the accumulation of the energy levels is described by the dispersion 
parameter (\ref{eq:Enu}) $\sigma_3=0.9082$.

We show comparison of the WKB result 
of Eq. (\ref{eq:q3-WKB}) with the exact expressions 
for $q_3$ at $h=1/2$
in Fig.~\ref{fig:N3q0} and Table~\ref{tab:WKB}. 
One can find that the
expression (\ref{eq:q3-WKB}) describes the excited eigenstates 
with good accuracy. In the case where
the eigenstates have smaller $q_3$
agreement becomes less accurate. Thus,
for the ground state with $iq_3=0.20526\ldots $ the accuracy of (\ref{eq:q3-WKB})
is $\sim 20\%$.
Obviously, in the region where the WKB expansion
is valid, \ie 
$|q_3^{1/3}|\gg |q_2^{1/2}|$, Eq.~(\ref{eq:q3-WKB}) can be systematically 
improved by including subleading WKB corrections.

\begin{table}[!ht]
\begin{center}
\begin{tabular}{|c||c|c|c|}
\hline
 $(\ell_1,\ell_2)$& $\left(q_3^{\rm \,exact}\right)^{1/3}$ & 
$\left(q_3^{\rm WKB}\right)^{1/3}$ & $-E_3/4$ 
\\
\hline
\hline

$(0,2)$ &  $0.590\,i$ & $0.684\,i$ & $-0.2472$ \\ \hline
$(2,2)$ & $0.358+0.621\,i$ & $0.395+0.684\,i$ & $-0.6910$ \\ \hline
$(4,2)$ & $0.749+0.649\,i$ & $0.790+0.684\,i$ & $-1.7080$ \\ \hline
$(6,2)$ & $1.150+0.664\,i$ & $1.186+0.684\,i$ & $-2.5847$ \\ \hline
$(8,2)$ & $1.551+0.672\,i$ & $1.581+0.684\,i$ & $-3.3073$ \\ \hline
$(10,2)$& $1.951+0.676\,i$ & $1.976+0.684\,i$ & $-3.9071$ \\ \hline

\end{tabular}
\end{center}
\caption[Comparison of the exact spectrum of $q_3^{1/3}$ at $h=1/2$ 
with the approximate
WKB expression (\ref{eq:q3-WKB}).]{Comparison 
of the exact spectrum of $q_3^{1/3}$ at $h=1/2$ 
with the approximate
WKB expression (\ref{eq:q3-WKB}). 
The last line defines the corresponding energy
$E_3(0;\ell_1,\ell_2)$.}
\lab{tab:WKB}
\end{table}

\subsection{Descendent states for $N=3$}

One also can notice in Figs. \ref{fig:prtraj} and \ref{fig:prtraj2}
that for odd $n_h$ we have states with 
$q_3=0$. For $n_h=0$ and $\nu_h=0$ it has the energy $E_3=0$, so
it is lower than (\ref{eq:N3-ground}). These states
are descendants 
of the states with two Reggeons
\ci{Lipatov:1998as,Bartels:1999yt,Bartels:2001hw,
Vacca:2000bk,Derkachov:2002wz}.
We constructed them in Section 3 using the $\oq{3}$ eigenfunction method.
The wave-functions of these states are described by 
(\ref{eq:Psib}),
(\ref{eq:BLV}),
(\ref{eq:Psiq0}) and
(\ref{eq:Psiq0b}).
These states have the same properties and the energy 
as the corresponding states with $N-1=2$ particles, 
$E_3(q_2,q_3=0)=E_2(q_2)$, with \ci{Balitsky:1978ic,Kuraev:1977fs}:
\begin{multline}
E_2(q_2)=4 \, \Re[\psi(1-h)+\psi(h)-2 \psi(1)]=\\
=8 \,\Re\left[\psi\left(\frac{1+|n_h|}{2}+i \nu_h\right)-\psi(1)\right]\,,
\lab{eq:EN=2}
\end{multline}
where  $\psi(x)=\frac{d}{dx} \ln \Gamma(x)$ and $q_2=-h(h-1)$.
Moreover, their wave-functions are built of 
the two-Reggeon states \ci{Vacca:2000bk,Derkachov:2002wz}
and the quasimomentum $\theta_3=0$.
Contrary to the states with $q_3 \ne 0$,
the states with $q_3=0$ (\ref{eq:BLV}) couple to a point-like
hadronic impact factors 
\ci{Engel:1997cg,Czyzewski:1996bv,Bartels:2003zu},
like the one for the $\gamma^* \to \eta_c$ transition. 

\subsection{Corrections to WKB}

The WKB formula for the lattice structure of the conformal charge $q_{3}$ 
was derived
in paper \ci{Derkachov:2002pb}. This formula tells us that
for $q_{3}\rightarrow \infty $
\begin{equation}
q_{3}^{1/3}=\frac{{\Gamma ^{3}(2/3)}}{2\pi }
{\cal Q}(\mybf n)\left[1+\frac{b}{\left|{\cal Q}(\mybf n)\right|^{2}}
-\left(\frac{b}{\left|{\cal Q}(\mybf n)\right|^{2}}\right)^{2}+
\sum _{k=3}^{\infty }{a_{k}
\left(\frac{b}{\left|{\cal Q}(\mybf n)\right|^{2}}\right)^{k}}\right]\,,
\lab{eq:korq3}
\end{equation}
where 
\begin{equation}
{\cal Q}(\mybf n)=\frac{1}{2}(l_{1}+l_{2})+i\frac{{\sqrt{3}}}{2}(l_{1}-l_{2})
=\sum _{k=1}^{3}{n_{k}e^{i\pi (2k-1)/3}}
\lab{eq:Qn}
\end{equation}
and $l_{1}$, $l_{2}$, ${\mybf n}=\{n_1,\ldots,n_N\}$
are integers, 
while the coefficient
\begin{equation}
b=\frac{3\sqrt{3}}{2\pi }{q_{2}}^{\ast}\,,
\lab{eq:korb}
\end{equation}
where {\it star} denotes complex conjugation.

After numerical calculations we have noticed that 
better agreement with the exact results is obtained for
\begin{equation}
b=\frac{3\sqrt{3}}{2\pi}
\left({q_{2}}^{\ast}-\frac{2}{3}\right)\,.
\lab{eq:bcoef}
\end{equation}
In order to show this,
we calculated the values of the conformal charge $q_{3}$ for 
$h=\frac{1+n_{h}}{2}$ for $n_h=0,1,\ldots,19$.
We evaluated numerically $q_3$ with  $\Im [q_3]=0$, \ie $a^{(r)}_k$,
and separately with $\Re [q_3]=0$, \ie $a^{(i)}_k$
where the superscript $(i)$ and $(r)$ refers to 
imaginary and real parts, respectively.
Then
we fitted expansion coefficients $a^{(r,i)}_{k}$ for large $|q_{3}|$
in the range $0\ldots 5000$ with high numerical precision.
In order to save space  in Table \ref{tab:q3coefs} we present
results only for $n_h=0,1,2$ and $3$.

\begin{table}[ht]
{
\scriptsize
\begin{center}
$\begin{array}{|c|c|r|r|r|r|r|} 
    \hline
n_h& \mbox{coef.} & k=3 \qquad & k=4 \qquad & k=5 \quad & k=6 \quad & k=7 \quad
 \\ \hline \hline
0
&a^{(r)}_k           & 0.509799695633 & -10.065761318 & -76.722084 & -1508.927 & -44580. \\ 
&a^{(i)}_k           & 3.490200304367 &  20.065761318 & 104.722084 &  -600.068 & -40411. \\
&a^{(r)}_k+a^{(i)}_k & 4.000000000000 &  10.000000000 &  28.000000 & -2108.995 & -84991. \\
    \hline
1
&a^{(r)}_k           &-2.585231705744 & -34.383865187 & -209.366287 & -1828.065 & -26404. \\
&a^{(i)}_k           & 6.585231705744 &  44.383865187 &  237.366287 &   340.913 & -16641. \\
&a^{(r)}_k+a^{(i)}_k & 4.000000000000 &  10.000000000 &   28.000000 & -1487.152 & -43046. \\
    \hline
2
&a^{(r)}_k           & 1.336317342408 &  1.2054728742 & -3.08256385 & -31.51219 & -196.87 \\
&a^{(i)}_k           & 2.663682657592 &  8.7945271258 & 31.08256385 & 110.89046 &  377.64 \\
&a^{(r)}_k+a^{(i)}_k & 4.000000000000 & 10.0000000000 & 28.00000000 &  79.37827 &  180.77 \\
    \hline
3
&a^{(r)}_k           & 2.250754858908 &  6.7220206127 & 22.76106772 & 79.325249 & 268.13 \\
&a^{(i)}_k           & 1.749245141092 &  3.2779793873 &  5.23893228 & -0.430149 & -75.81 \\
&a^{(r)}_k+a^{(i)}_k & 4.000000000000 & 10.0000000000 & 28.00000000 & 78.895100 & 192.32 \\
    \hline

\end{array}$
\end{center}}
\caption[The fitted coefficient to the series formula of $q_3^{1/3}$]
{The fitted coefficient to the series formula of $q_3^{1/3}$ 
(\ref{eq:korq3}) with  $n_h=0,1,2$ and $4$}
\lab{tab:q3coefs}
\end{table}

Coefficients $a_{k}$ for $k=0,\ldots ,2$ 
agree with formula (\ref{eq:korq3}),
\ie $a_{0}=1$, $a_{1}=1$ and  $a_{2}=-1$, but as we previously mentioned
in (\ref{eq:korb}) and (\ref{eq:bcoef}), the expansion parameter 
$b/\left|{\cal Q}(\mybf n)\right|^{2}$ is different.
This difference comes from the fact that the series formula (\ref{eq:korq3}) 
with (\ref{eq:korb}) is derived in the limit
$1 \ll |q_2^{1/2}| \ll |q_3^{1/3}|$.
Since in (\ref{eq:bcoef}) the value of $q_2^{\ast}$ is much bigger
then $2/3$ the value of the parameter $b$ from (\ref{eq:bcoef}) 
in the above limit goes to (\ref{eq:korb}).
One may suppose therefore that the factor $2/3$ 
in Eq. (\ref{eq:bcoef}) as subleading was omitted in
the derivation presented in
Ref. \ci{Derkachov:2002pb}.

Secondly, we see that the coefficients $a_k$ with $k>2$ 
start to depend on $n_h$. Thus, to describe the behaviour of 
$q_3^{1/3}$ properly, we have to introduce a second 
expansion parameter, for example $q_2$. 
We can also notice that 
for $k>2$ for  
a few first coefficients
$a_k^{(i)}+a_k^{(r)} \in \mathbb{Z}$ and this sum
does not depend on $n_h$.

Moreover, we can see that after the numerical fitting we obtain two different 
sets of the expansion coefficients, $\{a_k^{(r)}\}$ and $\{a_k^{(i)}\}$, 
defined in (\ref{eq:korq3}), for real  and 
imaginary  $q_3^{1/3}$, respectively. Thus, in order to describe
full-complex values of $q_3^{1/3}$ in terms of
the series (\ref{eq:korq3})
we have to use both sets of coefficients,
one for real and one for imaginary part of $q_3^{1/3}$.
Alternatively, we can perform expansion with two small parameters, \ie
$q^{\ast}_2/|{\cal Q}({\mybf n})|^2$ and $1/|{\cal Q}({\mybf n})|^2$.
Since the leading terms, \ie with $a_0$, $a_1$ and $a_2$, 
for real and imaginary $q_3^{1/3}$ are equal and
known analytically,
good approximation is obtained using
Eq. (\ref{eq:korq3})
with (\ref{eq:bcoef}) 
and neglecting higher order terms with $a_{k\ge 3}$.

\section{Summary}

In this work we have considered the scattering processes in the Regge limit
(\ref{eq:rlim}) where the compound reggeized gluon states, \ie Reggeons,
propagate in the $t-$channel and interact with each other.
We have performed calculations in the generalized leading logarithm
approximation (GLLA)
\ci{Bartels:1980pe,Kwiecinski:1980wb,Jaroszewicz:1980mq}, in which 
a number of Reggeons in the $t-$channel
is constant. 
We attempted to find a scattering amplitude of hadrons 
with multi-Reggeon exchange. However, a structure of reggeized gluon states
as well as their properties have turned out to be so reach, 
complicated and interesting that 
in this work we have focused on
description of the Reggeon state properties as well as
on analysing the spectra of the energy and integrals of motion.
The case for $N=2$ reggeized gluons was calculated in 
Ref.~\ci{Balitsky:1978ic,Kuraev:1977fs,Fadin:1975cb}.
Thus in this work we focused on $N=3$ Reggeon states 
and dependence of their spectrum on conformal Lorentz spin $n_h$
and scaling dimension $1 +2 \nu_h$ defined in (\ref{eq:hpar}).

In order to simplify the problem one applies the multi-colour limit 
 \ci{'tHooft:1973jz}, which makes the $N-$Reggeon system (\ref{eq:sepH})
$\SL(2,\mathbb{C})$ 
symmetric  (\ref{eq:trcoords}) and completely integrable. 
In this limit the equation for  the $N-$Reggeon wave-function
takes a form of Schr\"odinger equation (\ref{eq:Schr}) for
the non-compact XXX Heisenberg magnet model
of $\SL(2,\mathbb{C})$ spins $s$
\ci{Takhtajan:1979iv,Faddeev:1979gh,KBI}. 
Its Hamiltonian describes the nearest neighbour interaction
of the Reggeons \ci{Lipatov:1993yb,Faddeev:1994zg} 
propagating in the two-dimensional transverse-coordinates space 
(\ref{eq:trcoords}).
The system has 
a hidden cyclic and mirror permutation symmetry (\ref{eq:Psym}). 
It also possesses
the set of the $(N-1)$ of integrals of motion, which 
are eigenvalues of conformal charges  \ci{Derkachov:2001yn},
$\oq{k}$ and $\oqb{k}$,
(\ref{eq:qks0})--(\ref{eq:qks1}). Therefore, the operators of
conformal charges commute with each other and with
the Hamiltonian and they possess a common set of
the eigenstates.
However, for the $N=3$ case the multi-colour limit does not change
the Hamiltonian and its spectrum. We have the two integrals of motion
$\hat q_2$ and $\hat q_3$ and additionally constant two-dimensional
total momentum $p$.

Eigenvalues of the lowest conformal charge, $q_2$,
may be parameterized (\ref{eq:q2}) by the complex spins $s$
(\ref{eq:spins})
and the conformal weight $h$, where $h$ can be expressed by
the integer Lorentz spin $n_h$ and
the real parameter $\nu_h$ related to the scaling dimension (\ref{eq:hpar}).
Solving the eigenequation for $q_2$
we have derived ansatzes for $N-$Reggeons states with an arbitrary 
number of Reggeons $N$ as well as
arbitrary complex spins $s$.
Since the $N=3$ Reggeon ansatz separates variables,
the $q_3-$eigenequation can be rewritten as a 
differential equation of a Fuchsian type
with three singular points (\ref{eq:ees0z32}) \ci{Janik:1998xj}.
We have solved this equation by a series method.
Gluing solutions for different singular points,
(\ref{eq:u-zero}), (\ref{eq:u-one}) and (\ref{eq:u-inf}),
and taking care for normalization and single-valuedness
of the Reggeon wave-function
(\ref{eq:Psizz}) we have obtained 
the quantization conditions 
(\ref{eq:azao})--(\ref{eq:aoai})
for integrals of motion 
$(q_2,\wbar q_2,q_3,\wbar q_3)$
which we have solved numerically \ci{Derkachov:2002wz,Kotanski:2001iq}.
For $q_3=\wbar q_3=0$ 
the series solutions have a simple form
and we were able to resum them.
Thus, we have obtained analytical expressions for the 
three-Reggeon wave-functions with $q_3=0$
(\ref{eq:Psib})--(\ref{eq:Psiq0b}).

In this work we have calculated the behaviour of the 
$q_3-$spectrum
for the 
conformal Lorentz spins 
$n_h=0,1,2,3$
and  the scaling dimension $1+2 i \nu_h$.
Some results for $n_h>0$
were presented before in 
Ref.~\ci{DeVega:2001pu} for $n_h=1$ and
Refs.~\ci{Kotanski:2001iq} for $n_h=3$.
The quantized values of $q_3^{1/3}$ 
for given $n_h$ and fixed $\nu_h$
exhibit the WKB lattice structure (\ref{eq:q3-WKB}), which
for N=3 takes a form an equilateral-triangle lattice (\ref{eq:q3-WKB}),
Figs.\ \ref{fig:N3q0}-\ref{fig:prtraj2}.
The non-leading WKB corrections
move the quantized value of $q_3$ away from the lattice and
cause that
that the quantized values of $q_3$
lie outside a disk located around
the origin of the lattice, 
\ie near $q_3=0$.
However, for odd $n_h$ there exist
states with $q_3=0$. They are called 
descendent states because their wave-functions are 
effectively built of 
$N=2$ Reggeon states. 
The non-descendent state with the lowest energy
belongs to $n_h=0$ sector and its energy is positive
(\ref{eq:N3-ground}).
This state is double-degenerated and it
appears to be the nearest one to the origin of the $q_3^{1/3}$ lattice.
However, the ground state for $N=3$ is the descendent one with $n_h=1$
and energy $E_3=0$ (\ref{eq:EN=2}).
Having found the exact values of $q_3$
we are able to calculate corrections 
to the WKB approximation (\ref{eq:bcoef}),
Table \ref{tab:q3coefs}. 
These corrections differ from the corrections obtained earlier
in Ref.~\ci{Derkachov:2002pb}.
The difference seems to be caused by
using only one expansion parameter $\eta$ for
two various conformal charges, $q_2$ and $q_3$.
The obtained corrections are subleading to
the WKB approximation \ci{Derkachov:2002pb}
which is an expansion for large values of
conformal charges, \ie $1 \ll |q_2^{1/2}| \ll |q_3^{1/3}|$.

The above calculations
are of interest not only for perturbative QCD
but also to statistical physics 
as the $\SL(2,\mathbb{C})$ non-compact XXX Heisenberg spin magnet model
\ci{Takhtajan:1979iv,Faddeev:1979gh,KBI}.
This work opens the way for further studies
related to the multi-Reggeon states as well as
calculations of scattering amplitude for concrete processes.

\section*{Acknowledgements}
I would like to warmly thank to
Micha{\l} Prasza{\l}owicz
for fruitful discussions and help during writing this work.
I am very grateful to G.P. Korchemsky,
A.N.Manashov and S.\'E. Derkachov
whom I worked in an early state of this project.
I also thank to 
Jacek Wosiek
for illuminating discussions.
This work was supported by
KBN PB 2-P03B-43-24,
KBN PB 0349-P03-2004-27 and
KBN PB P03B-024-27(2004-2007).

\newpage

\newpage
\appendix
\newcommand{\LI}{\mbox{{\em L1}}}
\newcommand{\LII}{\mbox{{\em L2}}}
\newcommand{\LIII}{\mbox{{\em L3}}}
\newcommand{\NI}{\mbox{{\em N1}}}
\newcommand{\NII}{\mbox{{\em N2}}}
\newcommand{\NIII}{\mbox{{\em N3}}}
\newcommand{\M}{\mbox{\em M}}

\section{Conformal invariants and other variables}

Let us consider a difference of coordinates $(z_{1}-z_{2})$. It 
changes
under the $\SL(2,\mathbb{C})$ transformation (\ref{eq:trcoords}) as
\begin{equation}
 (z_{1}^{\prime }-z_{2}^{\prime })  = 
\left(\frac{az_{1}+b}{cz_{1}+d}-\frac{az_{2}+b}{cz_{2}+d}\right)=
(cz_{1}+d)^{-1}(cz_{2}+d)^{-1}(z_{1}-z_{2})\,.
\lab{eq:zdiff}
\end{equation}
One can see that during the transformation (\ref{eq:zdiff})
the factors $(cz_{i}+d)^{-1}$ appear
in front of the difference. 
These factors also exist in the transformation law of the wave-function
(\ref{eq:trpsiz0}). In order to cancel these factors one can construct
a fraction where the same additional factors appear 
in the denominator and numerator of the constructed fraction variable.

The fraction variable 
\begin{equation}
x=\frac{(z_{1}-z_{2})(z_{3}-z_{0})}{(z_{1}-z_{0})(z_{3}-z_{2})}\equiv (z_{1}z_{2}z_{3}z_{0})
\lab{eq:x}
\end{equation}
is invariant under the $\SL(2,\mathbb{C})$ transformations (\ref{eq:trcoords}).
There is only one independent invariant for four coordinates 
\ci{CFT,Staruszkiewicz93}.
One can see that the fractions coming from the $\SL(2,\mathbb{C})$
transformation (\ref{eq:zdiff}) cancel because
\begin{itemize}
\item the variable is a function of the coordinate differences 
(\ref{eq:zdiff}), 
\item the coordinates  
in the numerator and denominator of the variable $x$ are the same.
\end{itemize}
Therefore, simplifying the partial fractions we can obtain expression like $((az_{i}+b)(cz_{j}+d)-(az_{j}+b)(cz_{i}+d))$
which with making use of $ad-cb=1$ goes to $(z_{i}-z_{j})$. As we can
see, we have to build $\SL(2,\mathbb{C})-$invariants from differences of the coordinates.

It is easy to see that performing  permutations of coordinates we can construct
six different dependent invariants
\begin{equation}
\begin{array}{cccccc}
\displaystyle
 (z_{1}z_{2}z_{3}z_{0}) & = & x\mbox {,} & (z_{3}z_{2}z_{1}z_{0}) & = & 1/x\mbox {,}\\
 (z_{2}z_{3}z_{1}z_{0}) & = & 1/(1-x)\mbox {,} & (z_{1}z_{3}z_{2}z_{0}) & = & 1-x\mbox {,}\\
 (z_{3}z_{1}z_{2}z_{0}) & = & (x-1)/x\mbox {,} & (z_{2}z_{1}z_{3}z_{0}) & = & x/(x-1)\mbox {.}\end{array}\end{equation}

Let us take another product of $z_{ij}$
\begin{multline}
w^{\prime }  =  
\frac{(z_{1}^{\prime }-z_{2}^{\prime })}{
(z_{1}^{\prime }-z_{0}^{\prime })
(z_{2}^{\prime }-z_{0}^{\prime })}
=(cz_{0}+d)^{2}\frac{(z_{1}-z_{2})}{(z_{1}-z_{0})(z_{2}-z_{0})}=\\
=(cz_{0}+d)^{2}w\,.
\lab{eq:wdef}
\end{multline}
One can see that since in the denominator we have two additional $z_{0}$
variables after the transformation we 
obtain a multiplying factor $(cz_{0}+d)^{2}$
. 
Similarly for $w^h$ we obtain
\begin{equation}
w^{\prime h}=(cz_{0}+d)^{2h}w^{h}\,.
\end{equation}
Now one can compare this transformation to the transformation of
the $\SL(2,\mathbb{C})$ wave-function (\ref{eq:trpsiz0}).

For $N=3$ the variables $w=\frac{(z_3-z_2)}{(z_3-z_0)(z_2-z0)}$ 
and $x=\frac{(z_{1}-z_{2})(z_{3}-z_{0})}{(z_{1}-z_{0})(z_{3}-z_{2})}$ 
transform under the cycling permutation 
(\ref{eq:Psym}) as 
\begin{equation}
\begin{array}{c}
\displaystyle
(x-1)\rightarrow\frac{(-x)}{(x-1)}\rightarrow\frac{1}{(-x)}
\rightarrow(x-1)\,,\\
\displaystyle
(-x)\rightarrow\frac{1}{(x-1)}\rightarrow\frac{(x-1)}{(-x)}
\rightarrow(-x)\,,\\
\displaystyle
w\rightarrow w(x-1)\rightarrow w(-x)\rightarrow w
\end{array}
\lab{eq:wxcyc}
\end{equation}
while under the mirror permutation:
\begin{equation}
\begin{array}{c}
\displaystyle
(x-1)\rightarrow\frac{(x-1)}{(-x)}\rightarrow(x-1)\\
\displaystyle
(-x)\rightarrow\frac{1}{(-x)}\rightarrow(-x)\\
\displaystyle
w\rightarrow w\,x\rightarrow w
\end{array}
\lab{eq:wxmir}
\end{equation}

For higher $N$ we have more invariants. All of them can be constructed 
\ci{Lipatov:1998as}
from
\begin{equation}
x_{r}=\frac{(z_{r-1}-z_{r})(z_{r+1}-z_{0})}{(z_{r-1}-z_{0})(z_{r+1}-z_{r})};
\quad \prod _{r=1}^{N}x_{r}=(-1)^{N};
\quad \sum _{r=1}^{N}(-1)^{r}\prod _{k=r+1}^{N}x_{k}=0.
\lab{eq:xr}
\end{equation}
Variables $x_r$ are subject to the two conditions:
\begin{equation}
\qquad \prod _{r=1}^{N}x_{r}=(-1)^{N}
\lab{eq:xrc1}
\end{equation}
and 
\begin{equation}
\qquad \sum _{r=1}^{N}(-1)^{r}\prod _{k=r+1}^{N}x_{k}=0.
\lab{eq:xrc2}
\end{equation}
From (\ref{eq:xrc1}) we have
$x_{1}=(-1)^{N}/\prod _{r=2}^{N}x_{r}$. 
From (\ref{eq:xrc2}) we can
calculate 
\begin{equation}
x_{2}=\frac{\sum _{r=2}^{N}(-1)^{r}
\prod _{k=r+1}^{N}x_{k}}{\prod _{k=3}^{N}x_{k}}\,.
\end{equation}
Interchangeably,
one can derive from (\ref{eq:xrc1})
$x_{N}=(-1)^{N}/\prod _{r=1}^{N-1}x_{r}$ and next from
(\ref{eq:xrc2}) 
\begin{equation}
x_{N-1}=\frac{(-1)^{N}}{\sum _{r=1}^{N-1}(-1)^{r}
\prod _{k=r}^{N-2}x^{k}}\,.
\end{equation}
Thus, we see that we have for $N$ particles $N-2$ independent
invariants built of the particle coordinates $z_i$.

\section{Solutions for $N=3$ and $s=0$ around $x=0^{+}$}

The eigenequation for $\oq{3}$ (\ref{eq:qF})
is a differential equation of the third order 
so around each singular point, $x=0,1,\infty$, it
has three independent solutions. 
Around $x=0^{+}$ we have an indicial
equation 
\begin{equation}
(h-n-r)(r+n-1)(n+r)=0\,,
\lab{eq:indx0}
\end{equation} 
so its solutions are 
given by $r_{1}=h$, $r_{2}=1$ and $r_{3}=0$. 
As we can see we have two cases when 
$h\not\in\mathbb{Z}$
(one solution with $\mbox{Log}(x)$) and $h\in\mathbb{Z}$ (one solution
with $\mbox{Log}(x)$ and one solution with $\mbox{Log}^{2}(x)$). As we
will see below we also have to consider separately solutions with $q_{3}=0$.

Using the same method we can also found solutions around 
$x=1$ and $\infty$. However, to save the space we do not
present them.
 
\subsection{ Solutions for $q_3\ne0$ and $h\not\in\mathbb{Z}$ }

In the first case, \ie
for  $h\not\in\mathbb{Z}$ and  $q_3\ne0$
the solutions look as follows 
\begin{eqnarray}
\nonumber
u_{1}(x) & = & x^{r_{1}}\sum_{n=0}^{\infty}a_{n,r_{1}}x^{n}\,,\\
\nonumber
u_{2}(x) & = & x^{r_{2}}\sum_{n=0}^{\infty}a_{n,r_{2}}x^{n}\,,\\
u_{3}(x) & = & x^{r_{3}}\sum_{n=0}^{\infty}b_{n,r_{3}}x^{n}
+x^{r_{2}}\sum_{n=0}^{\infty}a_{n,r_{2}}x^{n}\mbox{Log}(x)\,,
\lab{eq:x0u123}
\end{eqnarray}
where $a_{0,r}$ is arbitrary (\eg equal to $1$)
\begin{equation}
a_{1,r}=\frac{(iq_{3}-(h-2r)(h-r)r)}{(h-1-r)r(1+r)}a_{0,r}
\end{equation}
and $m=n+r$
\begin{multline}
a_{n,r}=\frac{(h-m+1)(h-m+2)(m-2)}{(h-m)(m-1)m}a_{n-2,r}\\
+\frac{(iq_{3}-(1+h-m)(m-1)(h-2(m-1)))}{(h-m)(m-1)m}a_{n-1,r}\,,
\end{multline}
whereas $b_{0,r_{3}}=\frac{(h-1)}{iq_{3}}a_{0,r_{2}}$ 
and
$b_{1,r_{3}}$ is arbitrary.
One can notice that coefficient 
$b_{0,r_{3}}$ is well defined only for $q_{3}\ne0$.
Moreover,
\begin{equation}
b_{2,r_{3}}  = \frac{2+(h-3)h-iq_{3}}{2(2-h)}b_{1,r_{3}}
+\frac{3h-8}{2(2-h)}a_{1,r_{3}}+\frac{6+h(h-6)}{2(2-h)}a_{0,r_{3}}
\end{equation}
and
\begin{multline}
b_{n,r_{3}}  =  \frac{(h+1-m)(h-m+2)(m-2)}{(h-m)(m-1)m}b_{n-2,r_{3}}\\
+\frac{iq_{3}-(1+h-m)(m-1)(h-2(m-1))}{(h-m)(m-1)m}b_{n-1,r_{3}}\\
  +\frac{h^{2}+h(7-4m)+(m-2)(3m-4)}{(h-m)(m-1)m}a_{n-3,r_{3}}\\
-\frac{h^{2}-6h(m-1)+6(m-1)^{2}}{(h-m)(m-1)m}a_{n-2,r_{3}}
  -\frac{2m-3m^{2}+h(2m-1)}{(h-m)(m-1)m}a_{n-1,r_{3}}\,,
\end{multline}
where $m=r_{3}+n$.

\subsection{ Solution with $q_{3}=0$}

In this case $a_{0,r}$ defined in (\ref{eq:x0u123})
is arbitrary (\eg equal to $1$). 
For the first solution
with $r_{1}=h$ we have $a_{1,r_{1}}=0$ whereas for the third one $a_{1,r_{3}}$
is arbitrary (let us take $0$).
It turns out that we do not need $\mbox{Log}-$solutions.
The second solution is more complicated. One can derive the exact formula
for 
\begin{equation}
a_{n,r_{2}}=a_{0,r_{2}}\prod_{k=1}^{n}\frac{k-h}{k+1}
=\frac{\Gamma(1-h+n)}{\Gamma(1-h)\Gamma(n+2)}a_{0,r_{2}}
\end{equation}
and performing summations
\begin{eqnarray}
\nonumber
u_{1}(x) & = & x^{r_{1}}\sum_{n=0}^{\infty}a_{n,r_{1}}x^{n}=x^{h}a_{0,r_{1}}
\,,\\
\nonumber
u_{2}(x) & = & x^{r_{2}}\sum_{n=0}^{\infty}a_{n,r_{2}}x^{n}
=a_{0,r_{2}}x\sum_{n=0}^{\infty}
\frac{ \Gamma(1-h+n)}{\Gamma(1-h)\Gamma(n+2)}x^{n}=\\
\nonumber
&=&-a_{0,r_{2}}\frac{1}{h}((1-x)^{h}-1)\,,\\
u_{3}(x) & = & x^{r_{3}}\sum_{n=0}^{\infty}a_{n,r_{3}}x^{n}
=a_{0,r_{3}}\,.
\lab{eq:u123q0}
\end{eqnarray}
Gathering solutions (\ref{eq:u123q0}) we have
\begin{equation}
u(x)=A+B(-x)^{h}+C(x-1)^{h}
\end{equation}
where $A$, $B,$ $C$ are arbitrary. The above solution was presented
by Lipatov and Vacca in Refs.~\ci{Lipatov:1998as,Vacca:2000bk}.

\subsection{ Solution 
with $q_{3}\ne0$, $q_2=0$ and $h=1$ }

For $h=0$, \ie $q_2=0$, and $q_3 \ne 0$ 
we have a different set of solutions. Here
we have three solutions of the indicial equation 
(\ref{eq:indx0})
which are integer $r_{1}=1$,
$r_{2}=1$, $r_{3}=0$ so the solutions are given by
\begin{eqnarray}
u_{1}(x) & = & x^{r_{1}}\sum_{n=0}^{\infty}a_{n,r_{1}}x^{n}\,,\\
\nonumber
u_{2}(x) & = & x^{r_{2}}\sum_{n=0}^{\infty}b_{n,r_{2}}x^{n}
+x^{r_{1}}\sum_{n=0}^{\infty}a_{n,r_{1}}x^{n}\mbox{Log}(x)\,,\\
\nonumber
u_{3}(x) & = & x^{r_{3}}\sum_{n=0}^{\infty}c_{n,r_{3}}x^{n}
+2x^{r_{2}}\sum_{n=0}^{\infty}b_{n,r_{2}}x^{n}\mbox{Log}(x)
+x^{r_{1}}\sum_{n=0}^{\infty}a_{n,r_{1}}x^{n}\mbox{Log}^{2}(x)\,,
\end{eqnarray}
where $a_{0,r}$ is arbitrary (\eg equal $1$)
\begin{equation}
a_{1,r}=\frac{r+r^{2}(2r-3)}{r^{2}(1+r)-iq_{3}}a_{0,r}
\end{equation}
and $m=n+r$
\begin{equation}
a_{n,r}=-\frac{(m-3)(m-2)^{2}}{(m-1)^{2}m}a_{n-2,r}
+\frac{(m-2)(m-1)(2m-3)-iq_{3}}{(m-1)^{2}m}a_{n-1,r}\,,
\end{equation}
whereas 
\begin{equation}
b_{1,r_{2}}=\frac{1}{2}(-iq_{3}b_{0,r_{2}}+a_{0,r_{1}}-5a_{1,r_{1}})
\end{equation}
\begin{multline}
b_{n,r_{2}}  =  -\frac{(n-1)^{2}(n-2)}{n^{2}(n+1)}b_{n-2,r_{2}}
+\frac{(2m^{3}-3m^{2}+m-iq_{3})}{n^{2}(n+1)}b_{n-1,r_{2}}\\
+\frac{1+6n(n-1)}{n^{2}(n+1)}a_{n-1,r_{1}}+
  +\frac{(n-1)(5-3n)}{n^{2}(n+1)}a_{n-2,r_{1}}
-\frac{(2+3n)}{n(n+1)}a_{n,r_{1}}
\end{multline}
while $c_{0,r_{3}}=\frac{-2}{i q_{3}}a_{0,r_{1}}$, $c_{1,r_{3}}$
is arbitrary
and
\begin{equation}
c_{2,r_{3}}=-\frac{1}{2}iq_{3}c_{1,r_{3}}+b_{0,r_{2}}-5b_{1,r_{2}}
+3a_{0,r_{1}}-4a_{1,r_{1}}\,,
\end{equation}
\begin{multline}
c_{n,r_{3}} = -\frac{(n-2)^{2}(n-3)}{n(n-1)^{2}}c_{n-2,r_{3}}
+\frac{-iq_{3}-6+n(13+n(2n-9))}{n(n-1)^{2}}c_{n-1,r_{3}}\\
-\frac{2(n-2)(3n-8)}{n(n-1)^{2}}b_{n-2,r_{2}}
+\frac{26-36 n + 12 n^2}{n(n-1)^{2}}b_{n-1,r_{2}}\\
-\frac{6n^2-8n+2}{n(n-1)^{2}}b_{n,r_{2}}
+\frac{2(7-3n)}{n(n-1)^{2}}a_{n-3,r_{1}}
-\frac{6(3-2n)}{n(n-1)^{2}}a_{n-2,r_{1}}\\
+\frac{2(2-3n)}{n(n-1)^{2}}a_{n-1,r_{1}}\,.
\end{multline}

\subsection{ Solution with $q_2=q_{3}=0$ and $h=1$}

In this case $a_{0,r}$ is arbitrary (\eg equal to $1$). 
For the first solution
with $r_{1}=(h=1)$ we have $a_{1,r_{1}}=0$ and for the third one
$a_{1,r_{3}}$ with $r_{3}=0$ is arbitrary (let us take $0$). The
second solution is more complicated. We need $\mbox{Log}-$solutions
\begin{equation}
u_{2}(x)=x^{r_{2}}\sum_{n=0}^{\infty}b_{n,r_{2}}x^{n}
+x^{r_{1}}\sum_{n=0}^{\infty}a_{n,r_{1}}x^{n}\mbox{Log}(x)\,.
\end{equation}
Using recurrence  relations with $b_{1,r_{2}}=\frac{1}{2}a_{0,r_{1}}$
(and $a_{n,r_{1}}=0$ for $n>0$) 
\begin{eqnarray}
\nonumber
b_{2,r_{2}} & = & \frac{1}{3}b_{1,r_{2}}\,,\\
b_{n,r_{2}} & = & -\frac{(n-1)^{2}(n-2)}{n^{2}(n+1)}b_{n-2,r_{2}}
+\frac{(n-1)n(2n-1)}{n^{2}(n+1)}b_{n-1,r_{2}}
\end{eqnarray}
one can derive an exact formula for $b_{n,r_{2}}=b_{1,r_{2}}\frac{2}{(n+1)n}$
and performing summations with arbitrary $b_{0,r_{2}}(=0)$ we have
$x\sum_{n=0}^{\infty}b_{n,r_{2}}x^{n}=2b_{1,r_{2}}(x+\mbox{Log}(1-x)
-x\mbox{Log}(1-x))$, so that
\begin{eqnarray}
\nonumber
u_{1}(x) & = & x^{r_{1}}\sum_{n=0}^{\infty}a_{n,r_{1}}x^{n}=a_{0,r_{1}}
=xa_{0,r_{1}}\,,\\
\nonumber
u_{2}(x) & = & a_{0,r_{1}}(x+\mbox{Log}(1-x)-x\mbox{Log}(1-x)
+x\mbox{Log}(x))\,,\\
u_{3}(x) & = & x^{r_{3}}\sum_{n=0}^{\infty}a_{n,r_{3}}x^{n}
=a_{0,r_{3}}\,.
\lab{eq:u123h1}
\end{eqnarray}
Gathering solutions  (\ref{eq:u123h1}) we have
\begin{equation}
u(x)=A+B(-x)+C((x-1)\mbox{Log}(x-1)+(-x)\mbox{Log}(-x))\,,
\end{equation}
where $A$, $B,$ $C$ are arbitrary. See also Ref. \ci{DeVega:2001pu}.

\subsection{ Solution 
with $q_{3}\ne0$, $q_2=0$ and $h=0$}

For $h=0$, \ie $q_2=0$, and  arbitrary $q_3\ne 0$ 
we have three solutions 
\begin{eqnarray}
u_{1}(x) & = & x^{r_{1}}\sum_{n=0}^{\infty}a_{n,r_{1}}x^{n}\\
\nonumber
u_{2}(x) & = & x^{r_{2}}\sum_{n=0}^{\infty}b_{n,r_{2}}x^{n}
+x^{r_{1}}\sum_{n=0}^{\infty}a_{n,r_{1}}x^{n}\mbox{Log}(x)\\
\nonumber
u_{3}(x) & = & x^{r_{3}}\sum_{n=0}^{\infty}c_{n,r_{3}}x^{n}
+2x^{r_{2}}\sum_{n=0}^{\infty}b_{n,r_{2}}x^{n}\mbox{Log}(x)
+x^{r_{1}}\sum_{n=0}^{\infty}a_{n,r_{1}}x^{n}\mbox{Log}^{2}(x)
\nonumber
\end{eqnarray}
where $r_{1}=1$,
$r_{2}=0$ and $r_{3}=0$.
Here $a_{0,r}$ is arbitrary (\eg $1$)
\begin{equation}
a_{1,r}=\frac{-iq_{3}+2r^{2}}{r(1+r)^{2}}a_{0,r}
\end{equation}
and $m=n+r$
\begin{equation}
a_{n,r}=-\frac{(m-2)^{2}}{m^{2}}a_{n-2,r}
+\frac{-iq_{3}+2(m-1)^{3}}{(m-1)m^{2}}a_{n-1,r}\,,
\end{equation}
whereas $b_{0,r_{2}}=-\frac{1}{iq_{3}}a_{0,r_{1}}$ and $b_{1,r_{2}}$
is arbitrary while
\begin{equation}
b_{2,r_{2}}=\frac{1}{4}((2-iq_{3})b_{1,r_{2}}+6a_{0,r_{1}}-8a_{1,r_{1}})\,,
\end{equation}
\begin{multline}
b_{n,r_{2}}  =  -\frac{(n-2)^{2}}{n^{2}}b_{n-2,r_{2}}
+\frac{-iq_{3}+2(n-1)^{3}}{n^{2}(n-1)}b_{n-1,r_{2}}\\
-\frac{(n-2)(3n-4)}{n^{2}(n-1)}a_{n-3,r_{1}}
  +\frac{6(n-1)}{n^{2}}a_{n-2,r_{1}}
+\frac{2-3n}{n^{2}(n-1)}a_{n-1,r_{1}}\,.
\end{multline}
The coefficient
$c_{0,r_{3}}=\frac{2}{iq_{3}}(2a_{0,r_{1}}+b_{1,r_{2}})$ and
$c_{1,r_{3}}$ is arbitrary and
\begin{equation}
c_{2,r_{3}}=\frac{1}{4}(2-iq_{3})c_{1,r_{3}}+3b_{1,r_{2}}
-4b_{2,r_{2}}+3a_{0,r_{1}}-\frac{5}{2}a_{1,r_{1}}\,,
\end{equation}
\begin{multline}
c_{n,r_{3}}  =  -\frac{(n-2)^{2}}{n^{2}}c_{n-2,r_{3}}
+\frac{-iq_{3}+2(n-1)^{3}}{n^{2}(n-1)}c_{n-1,r_{3}}\\
-\frac{2(n-2)(3n-4)}{n^{2}(n-1)}b_{n-2,r_{2}}
+\frac{12(n-1)}{n^{2}}b_{n-1,r_{2}}+\frac{2(2-3n)}{n(n-1)}b_{n,r_{2}}\\
+\frac{2(5-3n)}{n^{2}(n-1)}a_{n-3,r_{1}}+\frac{12}{n^{2}}a_{n-2,r_{1}}
+\frac{2(1-3n)}{n^{2}(n-1)}a_{n-1,r_{1}}\,.
\end{multline}

\subsection{ Solution 
with $h=0$ and $q_2=q_{3}=0$}

In this case $a_{0,r}$ is arbitrary. For the first solution
with $r_{1}=1$ we have recurrence  relations with 
$a_{1,r_{1}}=\frac{1}{2}a_{0,r_{1}}$
\begin{eqnarray}
\nonumber
a_{2,r_{1}} & = & \frac{1}{3}a_{1,r_{1}}\,,\\
a_{n,r_{1}} & = & -\frac{(n-1)^{2}}{n+1}a_{n-2,r_{1}}
+\frac{2n^{2}}{n+1}a_{n-1,r_{1}}\,.
\end{eqnarray}
Summing series up we derive an exact formula for 
\begin{equation}
u_{1}(x)=x^{r_{1}}\sum_{n=0}^{\infty}a_{n,r_{1}}x^{n}
=x\sum_{n=0}^{\infty}x^{n}a_{0,r_{1}}\frac{1}{n+1}=-\mbox{Log}(1-x)\,.
\end{equation}

In the second solution, with $r_{2}=0$, we have arbitrary $a_{1,r_{2}}$
(let us take $0$) and solution is $u_{2}(x)=a_{0,r_{2}}$. The third
one is with $\mbox{Log}-$solutions 
\begin{equation}
u_{3}(x)=x^{r_{3}}\sum_{n=0}^{\infty}b_{n,r_{3}}x^{n}+x^{r_{2}}\sum_{n=0}^{\infty}a_{n,r_{2}}x^{n}\mbox{Log}(x)\,.
\end{equation}
Using recurrence relations with arbitrary $b_{0,r_{3}}$ 
\begin{eqnarray}
\nonumber
b_{2,r_{3}} & = & \frac{1}{2}b_{1,r_{3}}\,\\
b_{n,r_{3}} & = & -\frac{(n-2)^{3}}{n^{2}}b_{n-2,r_{3}}
+\frac{2(n-1)^{2}}{n^{2}}b_{n-1,r_{3}}
\end{eqnarray}
one can derive an exact formula for $b_{n,r_{3}}=b_{1,r_{3}}\frac{1}{n}$
and performing summations with arbitrary $b_{0,r_{3}}(=0)$ we have
$\sum_{n=0}^{\infty}b_{n,r_{3}}x^{n}=-b_{0,r_{3}}\mbox{Log}(1-x)$.
All these solutions look like 
\begin{eqnarray}
\lab{eq:u123h0}
u_{1}(x) & = & x^{r_{1}}\sum_{n=0}^{\infty}a_{n,r_{1}}x^{n}=-a_{1,r_{3}}\mbox{Log}(1-x)\,,\\
\nonumber
u_{2}(x) & = & x^{r_{2}}\sum_{n=0}^{\infty}a_{n,r_{2}}x^{n}=a_{0,r_{2}}\,,\\
\nonumber
u_{3}(x) & = & x^{r_{3}}\sum_{n=0}^{\infty}b_{n,r_{3}}x^{n}+a_{0,r_{2}}\mbox{Log}(x)
=-b_{0,r_{3}}\mbox{Log}(1-x)+a_{0,r_{2}}\mbox{Log}(x)\,.
\end{eqnarray}
Gathering solutions (\ref{eq:u123h0}) we obtain
\begin{equation}
u(x)=A+B\mbox{Log}(-x)+C\mbox{Log}(x-1)\,,
\end{equation}
where $A$, $B,$ $C$ are arbitrary.
See also Ref.\ \ci{DeVega:2001pu}.

\subsection{ Solution 
with $q_{3}\ne0$ and $h=2$}

For $h=2$ and $q_3\ne0$ 
we have three solutions 
\begin{eqnarray}
u_{1}(x) & = & x^{r_{1}}\sum_{n=0}^{\infty}a_{n,r_{1}}x^{n}\,,\\
\nonumber
u_{2}(x) & = & x^{r_{2}}\sum_{n=0}^{\infty}b_{n,r_{2}}x^{n}
+x^{r_{1}}\sum_{n=0}^{\infty}a_{n,r_{1}}x^{n}\mbox{Log}(x)\,,\\
\nonumber
u_{3}(x) & = & x^{r_{3}}\sum_{n=0}^{\infty}c_{n,r_{3}}x^{n}
+2x^{r_{2}}\sum_{n=0}^{\infty}b_{n,r_{2}}x^{n}\mbox{Log}(x)
+x^{r_{1}}\sum_{n=0}^{\infty}a_{n,r_{1}}x^{n}\mbox{Log}^{2}(x)
\end{eqnarray}
with $r_{1}=2$,
$r_{2}=1$, $r_{3}=0$. 
Here $a_{0,r_{1}}$ is arbitrary (\eg $1$)
\begin{equation}
a_{1,r_{1}}=\frac{-iq_{3}+2r(r-1)(r-2)}{r(r^{2}-1)}a_{0,r_{1}}
\end{equation}
and $m=n+r_{1}$\begin{equation}
a_{n,r_{1}}=-\frac{(m-3)(m-4)}{m(m-1)}a_{n-2,r_{1}}
+\frac{-iq_{3}+2(m-1)(m-2)(m-3)}{(m-2)(m-1)m}a_{n-1,r_{1}}\,,
\end{equation}
whereas $b_{0,r_{2}}=-\frac{2}{iq_{3}}a_{0,r_{1}}$ and $b_{1,r_{2}}$
is arbitrary while
\begin{equation}
b_{2,r_{2}}=\frac{1}{6}(-iq_{3}b_{1,r_{2}}+4a_{0,r_{1}}-11a_{1,r_{1}})\,,
\end{equation}
\begin{multline}
b_{n,r_{2}}  =  -\frac{(n-2)(n-3)}{n(n+1)}b_{n-2,r_{2}}
+\frac{-iq_{3}+2(n-2)(n-1)n}{n(n^{2}-1)}b_{n-1,r_{2}}\\
-\frac{11+3(n-4)n}{n(n^{2}-1)}a_{n-3,r_{1}}+
+\frac{6(n-2)n+4}{n(n^{2}-1)}a_{n-2,r_{1}}\\
+\frac{1-3n^2}{n(n^{2}-1)}a_{n-1,r_{1}}\,.
\end{multline}
Moreover, $c_{0,r_{3}}=\frac{2}{iq_{3}}b_{0,r_{2}}$ 
and $c_{1,r_{3}}=-\frac{4(b_{0,r_{2}}+b_{1,r_{2}})}{iq_{3}}
-\frac{6a_{0,r_{1}}}{iq_{3}}$
and $c_{2,r_{3}}$ is arbitrary while 
\begin{equation}
c_{3,r_{3}}=\frac{1}{6}(-iq_{3}c_{2,r_{3}}+2b_{0,r_{2}}
+8b_{1,r_{2}}-22b_{2,r_{2}}+12a_{0,r_{1}}-12a_{1,r_{1}})\,,
\end{equation}
\begin{multline}
c_{n,r_{3}}  =  -\frac{(n-3)(n-4)}{n(n-1)}c_{n-2,r_{3}}
+\frac{-iq_{3}+2(n-3)(n-2)(n-1)}{n(n-1)(n-2)}c_{n-1,r_{3}}\\
+\frac{2(3n(6-n)-26)}{n(n-1)(n-2)}b_{n-3,r_{2}}+
+\frac{4(11-12m+3m^{2})}{n(n-1)(n-2)}b_{n-2,r_{2}}\\
-\frac{2(2-6m+3m^{2})}{n(n-1)(n-2)}b_{n-1,r_{2}}
-\frac{6(n-3)}{n(n-1)(n-2)}a_{n-4,r_{1}}\\
+\frac{12}{n(n-1)}a_{n-3,r_{1}}
-\frac{6}{n(n-2)}a_{n-2,r_{1}}\,.
\end{multline}

\subsection{ Solution 
with $q_{3}=0$ and $h=2$}

In this case $a_{0,r}$ is arbitrary (\eg equal to $1$). 
For the first solution
with $r_{1}=2$ we have remaining coefficients $a_{n>0,r_{1}}=0$.
The second solution, with $r_{2}=1$, has an arbitrary $a_{1,r_{2}}$
(we can take it as $0$), $a_{2,r_{2}}=0$ and third one with arbitrary
$a_{1,r_{3}}$, $a_{2,r_{3}}$
(we also set them to $0$). All these
solutions look like 
\begin{eqnarray}
\nonumber
u_{1}(x) & = & x^{r_{1}}\sum_{n=0}^{\infty}a_{n,r_{1}}x^{n}
=x^{2}a_{0,r_{3}}\,,\\
\nonumber
u_{2}(x) & = & x^{r_{2}}\sum_{n=0}^{\infty}a_{n,r_{2}}x^{n}=xa_{0,r_{2}}\,,\\
u_{3}(x) & = & x^{r_{3}}\sum_{n=0}^{\infty}a_{n,r_{3}}x^{n}=a_{0,r_{3}}\,.
\lab{eq:u123h2}
\end{eqnarray}
Gathering solutions  (\ref{eq:u123h2}) we have
\begin{equation}
u(x)=A+B(-x)^{2}+C(x-1)^{2}\,,
\end{equation}
where $A$, $B,$ $C$ are arbitrary. As we can see this solution
corresponds to the solution with $q_{3}=0$ and arbitrary $h\not\in\{0,1\}$.
For other integer $h\not\in\{0,1\}$ we can observe the same correspondence.


\end{document}